\documentclass[%
reprint,
superscriptaddress,
 amsmath,amssymb,
 aps,
pra,
]{revtex4-2}

\usepackage{graphicx}
\usepackage{mathtools}

\DeclarePairedDelimiter\ket{\lvert}{\rangle}
\DeclarePairedDelimiterX\braket[2]{\langle}{\rangle}{#1\,\delimsize\vert\,\mathopen{}#2}

\newlength\figureheight 
\newlength\figurewidth 

\usepackage{dcolumn}% Align table columns on decimal point
\usepackage{bm}% bold math
\renewcommand{\figurename}{\textbf{Fig.}}

\usepackage{url}
\usepackage[colorlinks=true,allcolors=blue,bookmarks=false,pdfusetitle]{hyperref}
\usepackage{cleveref}

\begin{document}

\preprint{APS/123-QED}
\title{Bell correlations between momentum-entangled pairs of \textsuperscript{4}He\textsuperscript{*} atoms}

\author{Y. S. Athreya} 
\affiliation{% 
	Research School of Physics, Australian National University, Canberra 2601, Australia
}

\author{S. Kannan}
\affiliation{% 
	Research School of Physics, Australian National University, Canberra 2601, Australia
}
\author{X. T. Yan}
\affiliation{% 
	Research School of Physics, Australian National University, Canberra 2601, Australia
}
\author{R. J. Lewis-Swan}
\affiliation{% 
	Homer L. Dodge Department of Physics and Astronomy, The University of Oklahoma, Norman, Oklahoma 73019, USA }
\affiliation{% 
    Center for Quantum Research and Technology, The University of Oklahoma, Norman, Oklahoma 73019, USA
}
\author{K. V. Kheruntsyan}
\affiliation{% 
	School of Mathematics and Physics, University of Queensland, Brisbane, Queensland 4072, Australia
}
\author{A. G. Truscott}
\affiliation{% 
	Research School of Physics, Australian National University, Canberra 2601, Australia
}
\author{S. S. Hodgman}
\email{sean.hodgman@anu.edu.au}
\affiliation{% 
	Research School of Physics, Australian National University, Canberra 2601, Australia
}

\date{\today}% It is always \today, today,

\begin{abstract}
% \Com{[125 words.]}
Nonlocal entanglement between pair-correlated particles is a highly counter-intuitive aspect of quantum mechanics, where measurement on one particle can instantly affect the other, regardless of distance. 
While the rigorous Bell's inequality framework  has enabled the demonstration of such entanglement in photons and atomic internal states, no experiment has yet involved motional states of massive particles. Here we report the experimental observation of Bell correlations in motional states of momentum-entangled ultracold helium atoms. Momentum-entangled pairs are generated via $s$-wave collisions. Using a Rarity-Tapster interferometer and a Bell-test framework, we observe atom-atom correlations required for violation of a Bell inequality. This result shows the potential of ultracold atoms for fundamental tests of quantum mechanics and opens new avenues to studying gravitational effects in quantum states.

\end{abstract}

%\keywords{Suggested keywords}%Use show keys class option if keyword
\maketitle

%\tableofcontents

\section{\label{Introduction}Introduction}
Bell’s inequality serves as a fundamental test for distinguishing between classical local realism and the nonlocal correlations predicted by quantum mechanics \cite{Bell:book,aspect2004bellstheoremnaive}. Its violation is a cornerstone of quantum mechanics, directly challenging local hidden variable (LHV) theories and demonstrating the nonlocal nature of quantum entangled states. Such violations of Bell’s inequality have been experimentally observed in various systems, predominantly focusing on \emph{internal} degrees of freedom such as the polarization states of massless photons \cite{aspect1981ExperimentalTestsRealistic,aspect1982ExperimentalTestBells,LoopholeFreeBellPhotons2015,LoopholeFreeLocalRealism2015} and atomic spin states \cite{rowe2001ExperimentalViolationBells,hensen2015LoopholefreeBellInequality,schmied2016BellCorrelationsBoseEinstein,pezze2018QuantumMetrologyNonclassical,shinBell2019d}. 
These Bell tests involve measuring so-called Bell correlations -- a set of certain joint probability measurements on a pair of particles -- that provide compelling evidence supporting the nonlocal nature of quantum entanglement.

Extending Bell tests to \emph{external}, motional degrees of freedom -- particularly momentum-entanglement -- of massive particles offers
a deeper understanding of quantum nonlocality and its implications for the foundations of quantum mechanics. Momentum-entangled states of massive particles, for instance, enable fundamental experiments involving couplings to gravitational fields, thereby enabling tests of theories that seek to reconcile the currently incompatible frameworks of quantum mechanics and general relativity \cite{Born1938suggestion,penroseGravitys1996,Khrennikov2017,Howl2018gravity,Howl2019,Ashtekar2021}. However, experimental demonstrations of Bell inequality violations in motional states remain limited, with only photonic demonstrations to date \cite{rarityExperimental1990}.

Ultracold atomic systems, and particularly metastable helium ($^{4}\mathrm{He}^{*}$) atoms \cite{vassenCold2012}, have been proposed as promising systems for observing Bell nonlocality in momentum-entangled massive particle states \cite{lewis-swanProposal2015}. The high internal energy of metastable helium enables precise single-atom detection with high spatial and temporal resolution. Efforts towards this goal so far \cite{Kofler2012,Keller2014,dussarratTwoParticle2017} include demonstrating a matter-wave Rarity-Tapster configuration \cite{rarityExperimental1990} interferometer using colliding BECs of $^{4}\mathrm{He}^{*}$  \cite{thomasMatterwave2022} and showing control over the relative phases of momentum modes in atomic Bragg diffraction \cite{Leprince2024}. Additionally, Perrier \emph{et al.} \cite{Perrier2019} validated the quantum statistics of a two-mode squeezed vacuum state generated via atomic four-wave mixing in BEC collisions, either in free space \cite{perrinObservation2007,perrinAtomic2008,khakimovGhost2016a,thomasMatterwave2022}  or within an optical lattice potential \cite{Hilligsoe2005,Campbell2006,Bucker2011,Bonneau2013,Lopes2015,dussarratTwoParticle2017}. In the low-mode occupancy limit, such states approximate the archetypal Bell state that maximally violates Bell's inequality.  However, a demonstration of nonlocal behaviour in this system has remained elusive.

This work presents the first experimental observation of Bell correlations sufficient to demonstrate nonlocality in momentum-entangled pairs of atoms. By colliding two BECs of $^{4}\mathrm{He}^{*}$, we generate pairs of correlated atoms with opposite momenta via spontaneous  $s$-wave scattering. We implement the matter-wave analog of the Rarity-Tapster interferometric scheme and measure phase-sensitive momentum correlations between scattered atoms after passing through separate interferometric arms. Analyzing these momentum correlations within the framework of a Bell inequality test, we observe strong correlations that provide direct evidence of the nonlocal quantum nature of the system and are unable to be explained by a large class of LHV theories.

%-----------------------------------------------------------------------
\section{\label{Method}Method}

\begin{figure*}[t]
    \centering
    \includegraphics[scale=0.65]{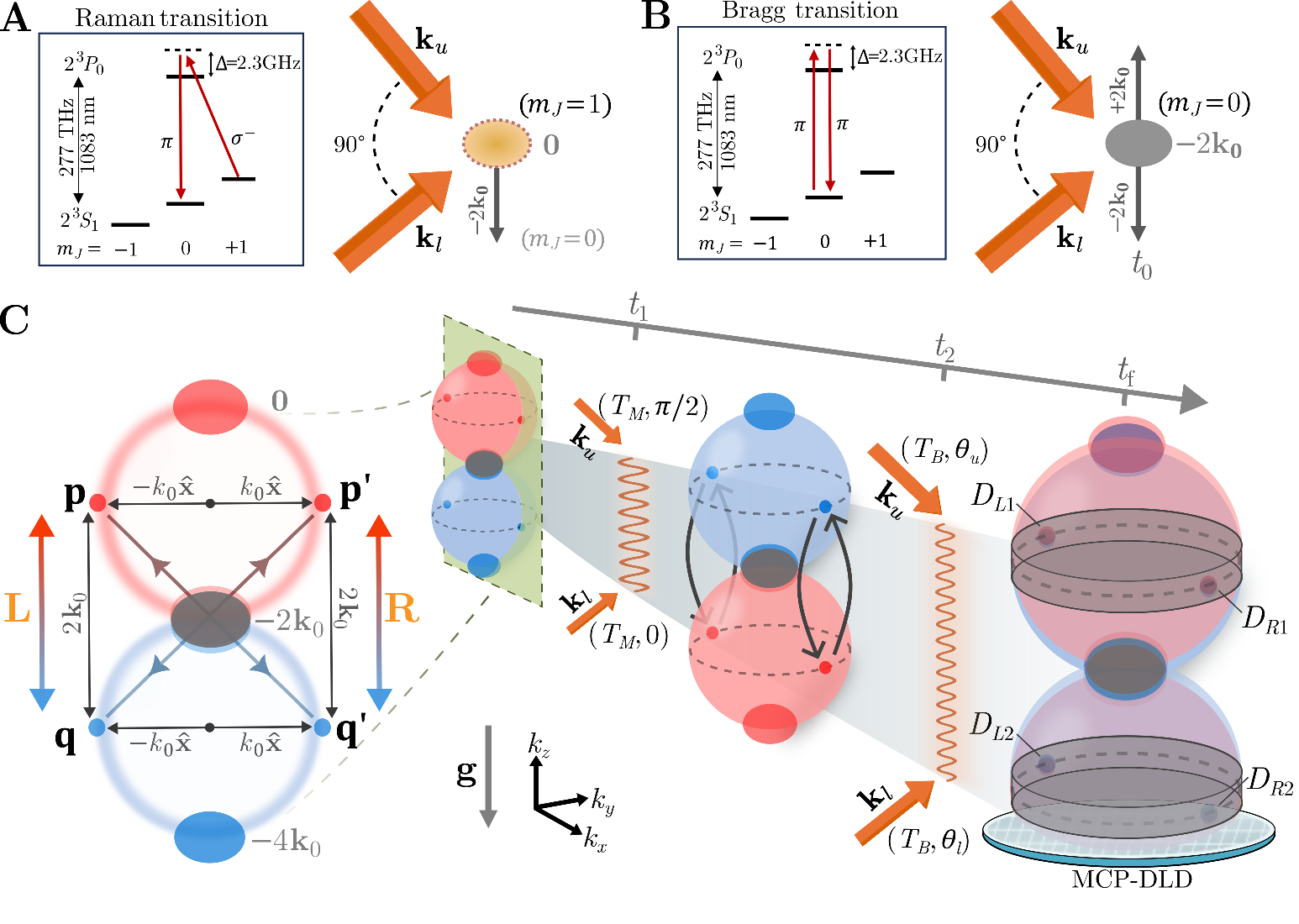}
    \caption{\textbf{Schematic of the experimental procedure in momentum space}. Orthogonal laser beams with wavevectors $\textbf{k}_u$ and $\textbf{k}_l$ (orange arrows) initially drive a two-photon Raman transition (\textbf{A}) to transfer $^4$He$^*$ atoms to the $m_J$=0 sublevel and impart momentum of $-2k_0\hat{\textbf{z}}$ in the direction of gravity ($\mathbf{g}$). (\textbf{B}) A Bragg transition at $t_0$ coherently splits the atoms into momentum modes: \textbf{0},$-2\textbf{k}_0$,$-4\textbf{k}_0$. The \textbf{0} and $-2\textbf{k}_0$ components (red) and $-2\textbf{k}_0$ and $-4\textbf{k}_0$ components (blue) collide to create two spherical $s$-wave scattering halos of entangled pairs of atoms (\textbf{C}). A Rarity-Tapster interferometer mixes atom pairs in momentum modes \{$\textbf{p},\textbf{p}^\prime$\} and \{$\textbf{q},\textbf{q}^\prime$\} from each halo, which are entangled through momentum conservation. Bragg transition pulses couple the atoms %\{$\textbf{p},\textbf{p}^\prime, \textbf{q},\textbf{q}^\prime $\} 
    in separate arms of the interferometer, denoted by $L$ for ($\textbf{p},\textbf{q}$) and $R$ for ($\textbf{p}^\prime,\textbf{q}^\prime$). At $t_1$ and $t_2$, we apply mirror and beamsplitter pulses of duration $T_M$ and $T_B$ and impart phases $\pi/2$ and $\phi_{L, R} = \theta_u-\theta_l$, respectively, equal to the phase difference between the upper ($\theta_u$) and lower ($\theta_l$) Bragg beams. After a fall-time $t_\text{f}\approx 0.416$ s the scattered atoms are detected on the MCP-DLD. Detection windows (grey-shaded annuli) centred around each halo's equatorial plane yield multi-particle correlations between \{$D_{L1}, D_{L2}, D_{R1}, D_{R2}$\} at the output of the interferometer. 
    }
    \label{fig:fig1}
\end{figure*}

Our experiment begins by creating a Bose-Einstein Condensate (BEC) of $\sim$10$^5$ helium atoms trapped in the $m_J=+1$ sub-level of the long-lived $2^{3}$S$_{1}$ metastable state \cite{hodgmanMetastable2009, vassenCold2012}. Figure \ref{fig:fig1} shows a schematic representation of the experimental procedure. Using a bi-planar quadrupole Ioffe configuration magnetic trap \cite{dallBose2007a}, we prepare our $^4$He$^*$ BEC in a harmonic potential with frequencies of $(\omega_x,\omega_y,\omega_z)/2\pi \approx (15,25,25)$ Hz. Following a rapid switch-off of the magnetic trap, we wait approximately 1.5 ms for the background magnetic field to stabilise to a uniform magnetic field $\textbf{B}_0\approx[0.5(\hat{\mathbf{x}}+\hat{\mathbf{z}})-0.8\hat{\mathbf{y}}]$ G.

The momentum-entanglement is created and manipulated via resonant two-photon Raman \cite{molerTheoretical1992} and Bragg \cite{kozumaCoherent1999} transitions using two orthogonal laser beams aligned along the $(\hat{\mathbf{x}}\pm \hat{\mathbf{z}})/\sqrt{2}$ directions. The beams are phase-locked and tuned to have optical frequencies far-detuned (to the blue) from the $2^{3}$S$_{1}$ $\rightarrow$ $2^{3}$P$_{0}$ transition by 2.3 GHz (Fig.~\ref{fig:fig1}, A and B), to minimize single-photon absorption.

We first transfer 90(5)$\%$ of the BEC atoms to the magnetically insensitive $m_J=0$ sub-level via a Raman pulse. This pulse also imparts a two-photon recoil momentum of $-2\hbar\textbf{k}_0$ in the $\hat{z}$-direction, where $\textbf{k}_0=k_0\hat{\mathbf{z}}$ and ${k}_0 = K/\sqrt{2}$ based on our beam geometry. Here $K=2\pi/\lambda$ is the wavenumber of the laser beam and $\lambda$=1083.19 nm denotes the wavelength of the incident laser beams. By transferring atoms to a magnetically insensitive state, we prevent momentum distortions that may occur due to stray magnetic fields in the vacuum chamber during the experiment.

We then coherently split the condensate of $m_J=0$ atoms, already moving with momentum $-2\hbar \textbf{k}_0$, by imparting $\pm 2\hbar\textbf{k}_0$ momentum to the atoms via a two-photon Bragg transition pulse \cite{guptaCoherent2001,khakimovGhost2016a} (Fig.~\ref{fig:fig1}B). This `collision pulse' splits the bulk of the condensate atoms into three momentum orders: \textbf{0}, $-2\hbar\textbf{k}_0$, $-4\hbar\textbf{k}_0$. As the momentum displaced condensates spatially separate, pairs of constituent atoms undergo $s$-wave collisions \cite{perrinObservation2007,hodgmanMetastable2009}, forming spherically symmetric halos of spontaneously scattered atom pairs in momentum space. Each halo is centred about the centre-of-mass (COM) momentum of the relevant pair of colliding condensates.  Since we split the condensate into three momentum orders, we observe the formation of two distinct scattering halos: one between \textbf{0} and $-2\hbar\textbf{k}_0$ (red), and another between $-2\hbar\textbf{k}_0$ and $-4\hbar\textbf{k}_0$ (blue), as illustrated in Fig.~\ref{fig:fig1}C.

This collision process creates momentum-entangled atom pairs analogous to the process of four-wave mixing in quantum optics \cite{perrinAtomic2008,rugwayCorrelations2011a,perrinObservation2007}, in which entangled photon pairs are generated through spontaneous parametric down-conversion \cite{couteauSpontaneous2018}. In this context, each of the halos can be characterized as an ensemble of two-mode squeezed vacuum states, with atom pairs occupying diametrically opposite momentum modes within the halo, in accordance with conservation of momentum and energy \cite{lewis-swanProposal2015,thomasMatterwave2022}. We operate our experiment in the low-gain regime with mode occupancies $\bar{n}\ll1$, i.e., a low average number of atoms in a scattered momentum mode \cite{hodgmanSolving2017}. 
Under such conditions, we analyse data with a single pair detected in the relevant output momentum ports of the Rarity-Tapster interferometer, which allows us to truncate and transform the halo’s mode-squeezed states into a prototypical Bell state for atoms \cite{lewis-swanProposal2015, thomasMatterwave2022, SOM}. The reduced and truncated initial state (i.e., the state of a quartet of scattering modes that form the initial state at the input ports of the interferometer) can be approximated to acquire the form of a prototypical Bell state \cite{lewis-swanProposal2015,thomasMatterwave2022}:  
\begin{equation}
\label{eq1}
\ket{\Psi} \approx \frac{1}{\sqrt{2}}(\ket{1}_{\textbf{p}}\ket{1}_{\textbf{p}^\prime}\ket{0}_{\textbf{q}}\ket{0}_{\textbf{q}^\prime} + \ket{0}_{\textbf{p}}\ket{0}_{\textbf{p}^\prime}\ket{1}_{\textbf{q}}\ket{1}_{\textbf{q}^\prime}),
\end{equation}
where (\textbf{p},$\textbf{p}^\prime$) and (\textbf{q},$\textbf{q}^\prime$) (illustrated in Fig.~\ref{fig:fig1}) correspond to correlated momentum modes in the top (red) and bottom (blue) halos, respectively, satisfying, $\textbf{p}+\textbf{p}^\prime=2\textbf{k}_0$,  $\textbf{q}+\textbf{q}^\prime=-2\textbf{k}_0$, 
$\textbf{p}-2\textbf{k}_0 =\textbf{q}$ and $\textbf{p}^\prime-2\textbf{k}_0 =\textbf{q}^\prime$.

We wait a separation time of 350~$\mu$s ($=t_1$) after the collision pulse ($t_0$) to apply a series of Bragg pulses - mirror (at $t_1$) and beamsplitter (at $t_2=t_1+350$ $\mu$s) pulses \cite{SOM,adamsAtom1994} - to selectively couple the momentum modes: (\textbf{p},$\textbf{p}^\prime$) and (\textbf{q},$\textbf{q}^\prime$), whose total paths through the interferometer are indistinguishable and interfere. The interference is manipulated through phases $\phi_L$ and $\phi_R$ imparted onto the momentum modes by the beamsplitter pulse. In our experimental setup, we are limited to equal phase settings $\phi_L = \phi_R$ controlled via the relative phase of a single pair of Bragg laser beams that we use to globally address all momentum modes within the halos.  We refer to this setup as a Rarity-Tapster type matter-wave interferometer \cite{thomasMatterwave2022,rarityExperimental1990} (Fig.~\ref{fig:fig1}C).

To detect atoms, we utilize a micro-channel plate (MCP) and a delay line detector (DLD) system located 848~mm below the trap that provides three-dimensional ($3$D) resolved detection with single atom resolution \cite{manningHanbury2010}. From spatial-temporal information the DLD records, we can reconstruct the velocity (i.e., momentum) distribution of the atoms, enabling us to measure multi-atom momentum correlations within our system \cite{hodgmanSolving2017}. We begin with a raw reconstruction of the $3$D momenta of the atoms and then proceed through several post-processing stages, involving coordinate transformations, filtering, and masking, to accurately determine the correlations between selected atomic momenta \cite{SOM}.

%----------------------------------------------------------------------
\section{\label{Results}Results and Discussion}

\begin{figure}[tbp]
    \centering
 \includegraphics[width=\linewidth, trim={0cm 0cm 0cm 0cm}, clip]{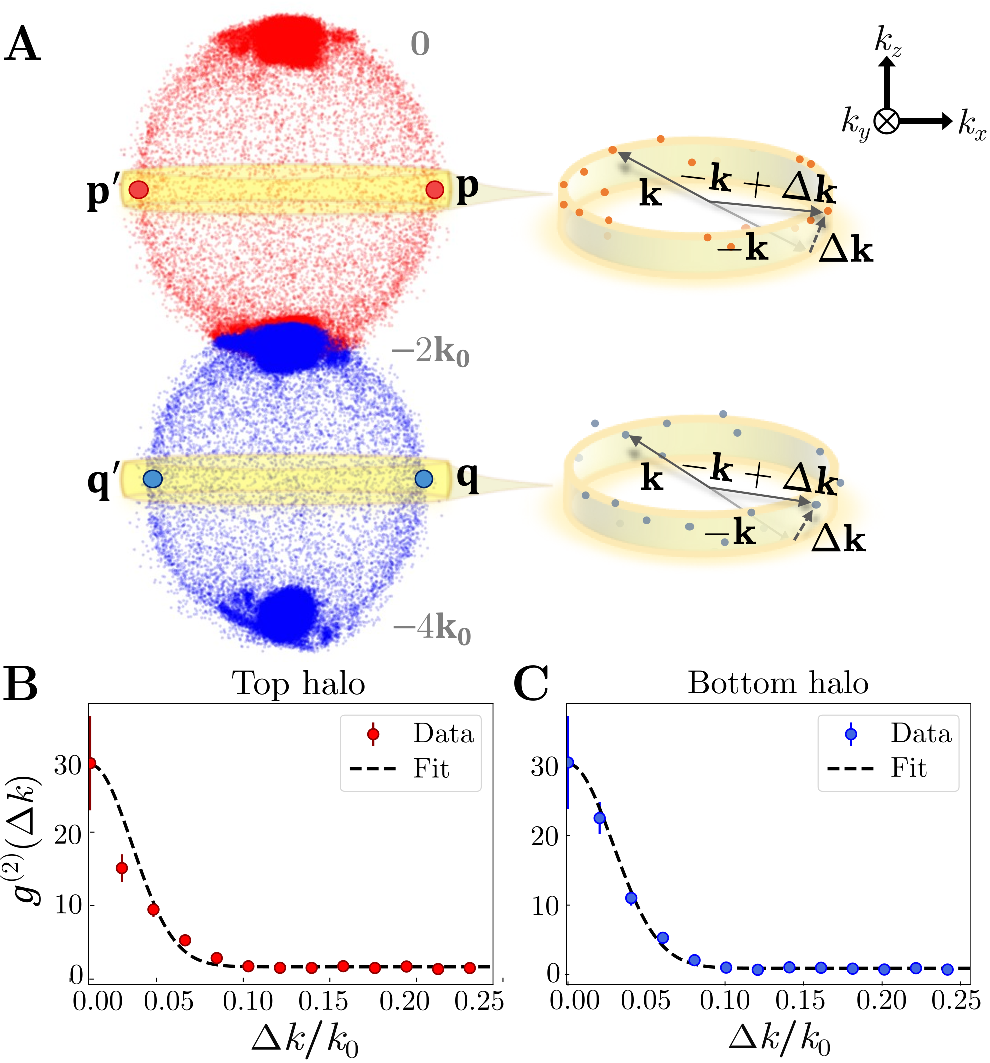}
    \caption{\textbf{Two-particle momentum correlations in scattering halos}. (\textbf{A}) Experimental data from 1000 shots showing the momentum distribution of the initial double $s$-wave scattering halo state. The yellow annulus about the equator of each halo depicts the chosen detection window range with a vertical range of $\pm4^{\circ}$ about the equator. (\textbf{B}), (\textbf{C}) Measured $g^{(2)}(\Delta k)$, with $\Delta k\equiv |\Delta\textbf{k}|$ and $k_0\equiv |\textbf{k}_0|$, for the `Top halo'- between $0$ and $-2\textbf{k}_0$ (red), and the `Bottom halo'- between $-2\textbf{k}_0$ and $-4\textbf{k}_0$ (blue), respectively. Error bars show the shot noise for each data point and solid lines are Gaussian fits to the data.}
    \label{fig:fig2}
\end{figure}

Forming the central basis in the analysis of our experiment is the measurement of two-particle momentum correlations \cite{hodgmanSolving2017} between atoms in opposite momentum modes {$\textbf{k}$, $-\textbf{k}+\Delta\textbf{k}$} of the scattering halos given by \cite{glauberQuantum1963b},
\begin{equation}
    g^{(2)}(\Delta\textbf{k}) \equiv g^{(2)}(\textbf{k},\textbf{-k}+\Delta\textbf{k}) = \frac{\sum_{\textbf{k}\in V}{\langle : \hat{n}_\textbf{k}\hat{n}_{\textbf{-k}+\Delta\textbf{k}}:\rangle}}{\sum_{\textbf{k}\in V}{\langle\hat{n}_\textbf{k}\rangle}\langle\hat{n}_{\textbf{-k}+\Delta\textbf{k}}\rangle},
\end{equation}
where $\hat{n}$ denotes the momentum-mode number operator and $V$ is the volume of the scattering halo occupied in momentum space (Fig.~\ref{fig:fig2}A). We describe this correlation function as the measurement of the joint probability of detecting atoms in the momenta \textbf{k},-\textbf{k}$+\Delta\textbf{k}$ divided by the product of their individual detection probabilities. Figures \ref{fig:fig2}B and C display the experimentally measured second-order correlation function $g^{(2)}(\Delta k)$ obtained from the pair of scattering halos, where $\Delta k \equiv|\Delta\textbf{k}|$. The high amplitude observed at $\Delta k=0$ indicates the generation of highly correlated atom pairs and is set by the mode occupancy $\bar{n}$ (following the relationship, $g^{(2)}(0) = 2+1/\bar{n}$ \cite{hodgmanSolving2017}), which represents the average number density of a scattering mode whose volume is set by the momentum correlation widths of the source condensate \cite{ogrenAtomatom2009}. We reach averaged amplitudes of $g^{(2)}(0)\sim 30$ for each of the halos (Fig.~\ref{fig:fig2}, B and C) corresponding to average mode occupancies of $\bar{n}\approx 0.035$, demonstrating 
% sufficient 
correlation amplitudes consistent with those required to violate a Bell inequality \cite{lewis-swanProposal2015,wasakBell2018}.

In the interferometer, the spatially separated atoms in the entangled momentum modes (\textbf{p}, \textbf{p}$^\prime$) and (\textbf{q}, \textbf{q}$^\prime$) must be made to overlap to observe significant multi-particle interference \cite{dussarratTwoParticle2017}. This interference depends on the phases applied by the beam splitter pulse on each separated arm of the interferometer (Fig.~\ref{fig:fig1}C). Successfully achieving this multi-particle interference with independently adjustable phases is essential to effectively demonstrate quantum non-locality 
% and to 
in the sense of a violation of the CHSH-Bell inequality \cite{clauserProposed1969}. We can observe this multi-particle interference at the output of our interferometer by measuring the joint probability distribution function or population correlations $P_{\textbf{k},\textbf{k}^\prime}$ between entangled momentum pairs (\textbf{p}, \textbf{p}$^\prime$) and (\textbf{q}, \textbf{q}$^\prime$). This is defined as $P_{\textbf{k},\textbf{k}^\prime} = \langle\hat{a}^\dagger_{\textbf{k}}\hat{a}^\dagger_{\textbf{k}^\prime}\hat{a}_{\textbf{k}^\prime}\hat{a}_{\textbf{k}}\rangle=\langle \hat{n}_\textbf{k}\hat{n}_{\textbf{k}^\prime}\rangle$ (where $\textbf{k} \in \{\textbf{p},\textbf{q}\}$ and $\textbf{k}^\prime \in \{\textbf{p}^\prime,\textbf{q}^\prime\}$) and measures correlations between joint-detection events at the outputs of the left ($L$) and right ($R$) arms of the interferometer. Taking the input of our interferometer as the momentum-entangled Bell state in Eq.\,\eqref{eq1} and treating the mirror and beamsplitter pulses of the interferometer as instantaneous linear transformations \cite{SOM}, we arrive at the joint probability distribution functions
\begin{equation}
P_{\textbf{p},\textbf{p}^\prime}=P_{\textbf{q},\textbf{q}^\prime} = \frac{1}{2}\sin^2(\frac{\phi_L+\phi_R}{2})=\frac{1}{2}\sin^2(\Phi/2) 
\label{eq:joint1}
\end{equation}
and
\begin{equation}
P_{\textbf{p},\textbf{q}^\prime}=P_{\textbf{q},\textbf{p}^\prime} = \frac{1}{2}\cos^2(\frac{\phi_L+\phi_R}{2})=\frac{1}{2}\cos^2(\Phi/2),
\label{eq:joint2}
\end{equation}
with respect to the output state. Interference of the scattered pairs is illustrated by the dependence of the joint probability distribution function on the 
% global
combined phase $\Phi = \phi_L + \phi_R$, where $\phi_L$ and $\phi_R$ represent the beamsplitter phases imparted on the momentum modes \{$\textbf{p},\textbf{q}$\} and \{$\textbf{p}^\prime,\textbf{q}^\prime$\}, respectively. 
Although our experimental configuration only allows for uniform control of the phases, i.e. $\phi_L = \phi_R$, Eqs.\,\eqref{eq:joint1} and \eqref{eq:joint2} demonstrate that it is still possible to use the global phase $\Phi=\phi_L+\phi_R$ (instead of the relative phase $\phi_L-\phi_R$ as in the original Rarity-Tapster scheme \cite{rarityExperimental1990,lewis-swanProposal2015}) to demonstrate entanglement and Bell correlations present in the initial state.

\begin{figure*}[t]
\centering
\includegraphics[width=\textwidth]{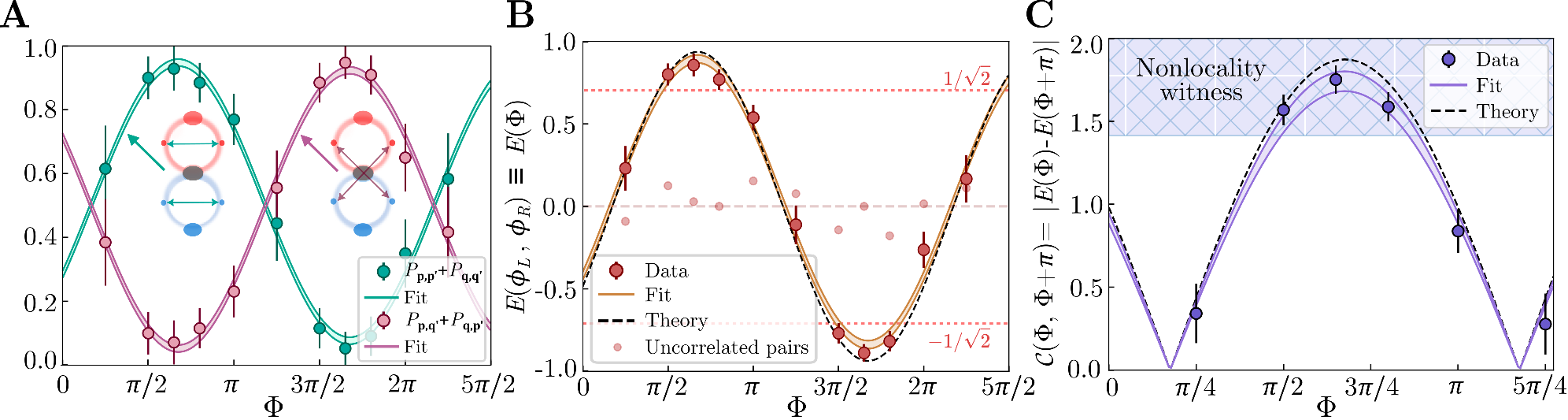}
    \caption{\textbf{Multi-particle interference and non-classical correlations}. Experimental data from over 35,000 shots. (\textbf{A}) Global phase ($\Phi$)-sensitive joint probability distribution function experimentally measured at the interferometer output as $\Phi$ is varied. Teal data points represent the measured values of ($P_{\textbf{p},\textbf{p}^\prime}$+$P_{\textbf{q},\textbf{q}^\prime}$), while magenta data points correspond to ($P_{\textbf{q},\textbf{p}^\prime}$+$P_{\textbf{p},\textbf{q}^\prime}$). (\textbf{B}) The Bell correlation function $E(\Phi)$ with a sinusoidal fit following Eq.\,\eqref{eq:eq5a}, where we obtain an amplitude of $A=0.86(3)$ and phase-offset $\delta=1.02(4)$. The light-coloured data points around $E=0$ indicate uncorrelated atom pairs passing through the interferometer, demonstrating expected near-zero Bell correlation amplitudes \cite{SOM}. (\textbf{C}) A nonlocality witness $\mathcal{C}$ constructed from values of the Bell correlation function $E$ at complimentary phase settings. Data points located within the shaded region ($>\sqrt{2}$) indicate the presence of Bell correlations strong enough to exclude a class of LHV theories, with an observed maximum violation of $\sim3.9\sigma$ at $\mathcal{C}(\Phi,\Phi+\pi)=1.752\pm0.085$ when $\Phi=13\pi/20$. All error bars correspond to standard deviation estimates using a binomial proportional estimator \cite{SOM} with solid lines being %sine-squared ((a)) or cosine ((b),(c)) 
    fits to the data. The theoretical prediction (black-dashed line) is Eq.\,\eqref{eq:eq5a}. 
    }
    \label{fig:fig3}
\end{figure*}

Figure \ref{fig:fig3}A displays the experimental joint probability distribution functions measured at the output of the interferometer as we vary the global phase $\Phi$. We observe strong out-of-phase oscillations of the joint probabilities consistent with the predictions of the ideal Bell state (Eqs.\,\eqref{eq:joint1} and \eqref{eq:joint2}), confirming the existence of multi-particle interference in our setup.

From the joint probability distribution functions, we can obtain the Bell correlation function
\begin{equation}
    E(\Phi) = \frac{P_{\textbf{p},\textbf{p}^\prime}+P_{\textbf{q},\textbf{q}^\prime}-P_{\textbf{p},\textbf{q}^\prime}-P_{\textbf{q},\textbf{p}^\prime}}{P_{\textbf{p},\textbf{p}^\prime}+P_{\textbf{q},\textbf{q}^\prime}+P_{\textbf{p},\textbf{q}^\prime}+P_{\textbf{q},\textbf{p}^\prime}} ,
    \label{eq:eq5}
\end{equation}
which is expected to have the general form \cite{Clauser_Shimony_review_1978}
\begin{equation}
    E(\Phi) = - A\cos(\Phi+\delta).
    \label{eq:eq5a}
\end{equation}
  The ideal Bell state (Eq.\,\eqref{eq1}) would yield an amplitude of $A=1$, following from Eqs.\,\eqref{eq:joint1} and \eqref{eq:joint2}. However, contributions from higher-order Fock states in the entangled pair generation process lead to a reduced amplitude $A=(1+\bar{n})/(1+3\bar{n})$ expressed in terms of the average mode occupancy $\bar{n}$ \cite{lewis-swanProposal2015,lewis-swanUltracold2016,wasakBell2018}. The additional phase offset $\delta$ in Eq.\,\eqref{eq:eq5a} accounts for details in the precise implementation of the Rarity-Tapster interferometer, such as path length differences between the two Bragg beams and phase drifts experienced by the scattered particles as they follow free-fall trajectories in a uniform gravitational potential.

The Bell correlation function $E(\Phi)$  constructed from the experimentally detected joint probabilities is shown in Fig.~\ref{fig:fig3}B. We find remarkable agreement with both a sinusoidal fit function, consistent with the expectation (Eq.\,\eqref{eq:eq5a}), as well as theory predictions (represented by the black-dotted line) with $A=(1+\bar{n})/(1+3\bar{n})$ computed from the experimentally obtained average mode occupancy. 

To verify that the observed Bell correlation function  demonstrates non-classical non-local behaviour, we construct the following nonlocality criterion \cite{SOM} using the values that $E(\Phi)$ takes at complementary global phase settings,
\begin{equation}
    \mathcal{C}(\Phi,\Phi+\pi)=|E(\Phi)-E(\Phi+\pi)|\leq \sqrt{2}.
    \label{eq:eq6}
\end{equation}
Violation of this bound witnesses not just that the quantum system is entangled but also rules out a wide range of LHV theories where one subsystem yields binary (classical-like) outcomes, while the other subsystem produces (quantum-like) outcomes that behave like components of a vector \cite{shinBell2019d,wisemanSteering2007a,cavalcantiExperimental2009a, SOM}. We plot the LHS of Eq.\,\eqref{eq:eq6} in Fig.~\ref{fig:fig3}C using the experimentally constructed values of $E(\Phi)$ (markers), a sinusoidal fit of $E(\Phi)$ (solid mauve line) and theory predictions (black-dotted line). The strong oscillations in the Bell correlation function lead to an observed maximum violation of $\mathcal{C}(\Phi,\Phi+\pi)$ = $1.752\pm 0.085 >\sqrt{2}$ (i.e., a violation of about $\sim3.9\sigma$) at $\Phi = 13\pi/20$.

The violation of the inequality (Eq.\,\eqref{eq:eq6}) is dependent on the Bell correlation function $E(\Phi)$ displaying a sufficiently large difference between the maximal and minimal correlation, as well as following a particular functional form. In the context of Eq.\,\eqref{eq:eq5a}, this reduces to a requirement that the oscillation amplitude exceeds $A > 1/\sqrt{2}$. Such a requirement on $A$ is also shared by more stringent tests of nonlocality, including the celebrated CHSH-Bell inequality \cite{clauserProposed1969,Clauser_Shimony_review_1978}. However, in the context of Bell nonlocality an equally important aspect that leads to the incompatibility of quantum mechanics with LHV theories is the sinusoidal variation of the Bell correlation function \cite{aspect2004bellstheoremnaive}. In Fig.~\ref{fig:fig3}B we demonstrate that our experimental observations feature both key elements by fitting Eq.\,\eqref{eq:eq5a} to our experiment observations, which yields a fitted value of $A=0.86(3)$. This signals the detection of Bell correlations in our experiment with the potential to violate the CHSH-Bell inequality upon the implementation of independent phase settings in the spatially separate regions of $L$ and $R$.

%-----------------------------------------------------------------------
\section{\label{Conclusion}Conclusion}
In conclusion, we generate momentum-entangled pairs of massive particles by colliding atomic BECs to form dual $s$-wave collision halos. Using Bragg beams, we coherently manipulate the momentum states of the scattered atoms, allowing for selective coupling of desired atom pairs in momentum states (\textbf{p},\textbf{p}$^\prime$,\textbf{q},\textbf{q}$^\prime$) and imparting an arbitrary global phase onto them.

We have observed two-particle correlations dependent on the global interferometric phase $\Phi$ and having a sufficiently strong amplitude $A$ that signal the detection of nonclassical, nonlocal Bell correlations in our experiment.
This demonstration could be extended by adding independent phase settings in separate regions of the scattering halos to test a CHSH-Bell inequality violation \cite{clauserProposed1969} using momentum-entangled states of atoms, an even stronger bound on nonlocality.

The future development of this scheme could potentially involve generating momentum-entanglement between isotopes of helium—specifically, $^3$He$^*$ and $^4$He$^*$ \cite{yanProposal2024a}. This entanglement between momentum states of atoms with different masses would offer a suitable basis for the testing of the weak equivalence principle with quantum test masses \cite{geigerProposal2018a}. Furthermore, such a platform for entangling massive particles could be useful in examining decoherence theories in quantum systems influenced by gravitational field interactions \cite{alvarezQuantum1989,penroseGravitys1996,zychQuantum2011,voweLightpulseAtomInterferometric2022}, and enhance our understanding of the relation between quantum theory and gravity, as described by general relativity. 

Our findings not only establish a new platform for testing the fundamental principles of quantum mechanics but also open avenues for exploring quantum information protocols that leverage motional entanglement \cite{braunsteinQuantumInformationContinuous2005,OSullivanHalePixelEntanglement2005,WalbornQKDSpatialQubits2006}. Demonstrating and controlling momentum-entanglement in ultracold atomic systems holds promise for advancing quantum technologies such as quantum sensing and quantum imaging through sub-shot noise atom interferometry \cite{anders2021MomentumEntanglementAtom,fadelQuantumMetrologyContinuousvariable2024}.

\begin{acknowledgements}
The authors would like to thank K.F. Thomas for technical assistance with the early stages of the experiment, and S.A. Haine for helpful discussions. This work was supported through the Australian Research Council (ARC) Discovery Projects, Grant No. DP190103021, DP240101346, DP240101441 and DP240101033.
S.S.H. was supported by the Australian Research Council Future Fellowship Grant No. FT220100670.
S.K. was supported by an Australian Government Research Training Program scholarship. 
\end{acknowledgements}

\bibliography{bell_main}% Produces the bibliography via BibTeX.

\begin{thebibliography}{67}%
\makeatletter
\providecommand \@ifxundefined [1]{%
 \@ifx{#1\undefined}
}%
\providecommand \@ifnum [1]{%
 \ifnum #1\expandafter \@firstoftwo
 \else \expandafter \@secondoftwo
 \fi
}%
\providecommand \@ifx [1]{%
 \ifx #1\expandafter \@firstoftwo
 \else \expandafter \@secondoftwo
 \fi
}%
\providecommand \natexlab [1]{#1}%
\providecommand \enquote  [1]{``#1''}%
\providecommand \bibnamefont  [1]{#1}%
\providecommand \bibfnamefont [1]{#1}%
\providecommand \citenamefont [1]{#1}%
\providecommand \href@noop [0]{\@secondoftwo}%
\providecommand \href [0]{\begingroup \@sanitize@url \@href}%
\providecommand \@href[1]{\@@startlink{#1}\@@href}%
\providecommand \@@href[1]{\endgroup#1\@@endlink}%
\providecommand \@sanitize@url [0]{\catcode `\\12\catcode `\$12\catcode `\&12\catcode `\#12\catcode `\^12\catcode `\_12\catcode `\%12\relax}%
\providecommand \@@startlink[1]{}%
\providecommand \@@endlink[0]{}%
\providecommand \url  [0]{\begingroup\@sanitize@url \@url }%
\providecommand \@url [1]{\endgroup\@href {#1}{\urlprefix }}%
\providecommand \urlprefix  [0]{URL }%
\providecommand \Eprint [0]{\href }%
\providecommand \doibase [0]{https://doi.org/}%
\providecommand \selectlanguage [0]{\@gobble}%
\providecommand \bibinfo  [0]{\@secondoftwo}%
\providecommand \bibfield  [0]{\@secondoftwo}%
\providecommand \translation [1]{[#1]}%
\providecommand \BibitemOpen [0]{}%
\providecommand \bibitemStop [0]{}%
\providecommand \bibitemNoStop [0]{.\EOS\space}%
\providecommand \EOS [0]{\spacefactor3000\relax}%
\providecommand \BibitemShut  [1]{\csname bibitem#1\endcsname}%
\let\auto@bib@innerbib\@empty
%</preamble>
\bibitem [{\citenamefont {Bell}(1987)}]{Bell:book}%
  \BibitemOpen
  \bibfield  {author} {\bibinfo {author} {\bibfnamefont {J.~S.}\ \bibnamefont {Bell}},\ }\href@noop {} {\emph {\bibinfo {title} {Speakable and Unspeakable in Quantum Mechanics}}}\ (\bibinfo  {publisher} {Cambridge Univ. Press},\ \bibinfo {address} {Cambridge},\ \bibinfo {year} {1987})\BibitemShut {NoStop}%
\bibitem [{\citenamefont {Aspect}(2004)}]{aspect2004bellstheoremnaive}%
  \BibitemOpen
  \bibfield  {author} {\bibinfo {author} {\bibfnamefont {A.}~\bibnamefont {Aspect}},\ }\bibfield  {title} {\bibinfo {title} {Bell's theorem: The naive view of an experimentalist},\ }\href {https://doi.org/10.48550/arXiv.quant-ph/0402001} {\bibfield  {journal} {\bibinfo  {journal} {arXiv:quant-ph/0402001}\ } (\bibinfo {year} {2004})}\BibitemShut {NoStop}%
\bibitem [{\citenamefont {Aspect}\ \emph {et~al.}(1981)\citenamefont {Aspect}, \citenamefont {Grangier},\ and\ \citenamefont {Roger}}]{aspect1981ExperimentalTestsRealistic}%
  \BibitemOpen
  \bibfield  {author} {\bibinfo {author} {\bibfnamefont {A.}~\bibnamefont {Aspect}}, \bibinfo {author} {\bibfnamefont {P.}~\bibnamefont {Grangier}},\ and\ \bibinfo {author} {\bibfnamefont {G.}~\bibnamefont {Roger}},\ }\bibfield  {title} {\bibinfo {title} {Experimental {{Tests}} of {{Realistic Local Theories}} via {{Bell}}'s {{Theorem}}},\ }\href {https://doi.org/10.1103/PhysRevLett.47.460} {\bibfield  {journal} {\bibinfo  {journal} {Phys. Rev. Lett.}\ }\textbf {\bibinfo {volume} {47}},\ \bibinfo {pages} {460} (\bibinfo {year} {1981})}\BibitemShut {NoStop}%
\bibitem [{\citenamefont {Aspect}\ \emph {et~al.}(1982)\citenamefont {Aspect}, \citenamefont {Dalibard},\ and\ \citenamefont {Roger}}]{aspect1982ExperimentalTestBells}%
  \BibitemOpen
  \bibfield  {author} {\bibinfo {author} {\bibfnamefont {A.}~\bibnamefont {Aspect}}, \bibinfo {author} {\bibfnamefont {J.}~\bibnamefont {Dalibard}},\ and\ \bibinfo {author} {\bibfnamefont {G.}~\bibnamefont {Roger}},\ }\bibfield  {title} {\bibinfo {title} {Experimental {{Test}} of {{Bell}}'s {{Inequalities Using Time- Varying Analyzers}}},\ }\href {https://doi.org/10.1103/PhysRevLett.49.1804} {\bibfield  {journal} {\bibinfo  {journal} {Phys. Rev. Lett.}\ }\textbf {\bibinfo {volume} {49}},\ \bibinfo {pages} {1804} (\bibinfo {year} {1982})}\BibitemShut {NoStop}%
\bibitem [{\citenamefont {Giustina}\ \emph {et~al.}(2015)\citenamefont {Giustina}, \citenamefont {Versteegh}, \citenamefont {Wengerowsky}, \citenamefont {Handsteiner}, \citenamefont {Hochrainer}, \citenamefont {Phelan}, \citenamefont {Steinlechner}, \citenamefont {Kofler}, \citenamefont {Larsson}, \citenamefont {Abell\'an}, \citenamefont {Amaya}, \citenamefont {Pruneri}, \citenamefont {Mitchell}, \citenamefont {Beyer}, \citenamefont {Gerrits}, \citenamefont {Lita}, \citenamefont {Shalm}, \citenamefont {Nam}, \citenamefont {Scheidl}, \citenamefont {Ursin}, \citenamefont {Wittmann},\ and\ \citenamefont {Zeilinger}}]{LoopholeFreeBellPhotons2015}%
  \BibitemOpen
  \bibfield  {author} {\bibinfo {author} {\bibfnamefont {M.}~\bibnamefont {Giustina}}, \bibinfo {author} {\bibfnamefont {M.~A.~M.}\ \bibnamefont {Versteegh}}, \bibinfo {author} {\bibfnamefont {S.}~\bibnamefont {Wengerowsky}}, \bibinfo {author} {\bibfnamefont {J.}~\bibnamefont {Handsteiner}}, \bibinfo {author} {\bibfnamefont {A.}~\bibnamefont {Hochrainer}}, \bibinfo {author} {\bibfnamefont {K.}~\bibnamefont {Phelan}}, \bibinfo {author} {\bibfnamefont {F.}~\bibnamefont {Steinlechner}}, \bibinfo {author} {\bibfnamefont {J.}~\bibnamefont {Kofler}}, \bibinfo {author} {\bibfnamefont {J.-A.}\ \bibnamefont {Larsson}}, \bibinfo {author} {\bibfnamefont {C.}~\bibnamefont {Abell\'an}}, \bibinfo {author} {\bibfnamefont {W.}~\bibnamefont {Amaya}}, \bibinfo {author} {\bibfnamefont {V.}~\bibnamefont {Pruneri}}, \bibinfo {author} {\bibfnamefont {M.~W.}\ \bibnamefont {Mitchell}}, \bibinfo {author} {\bibfnamefont {J.}~\bibnamefont {Beyer}}, \bibinfo {author} {\bibfnamefont {T.}~\bibnamefont {Gerrits}}, \bibinfo {author}
  {\bibfnamefont {A.~E.}\ \bibnamefont {Lita}}, \bibinfo {author} {\bibfnamefont {L.~K.}\ \bibnamefont {Shalm}}, \bibinfo {author} {\bibfnamefont {S.~W.}\ \bibnamefont {Nam}}, \bibinfo {author} {\bibfnamefont {T.}~\bibnamefont {Scheidl}}, \bibinfo {author} {\bibfnamefont {R.}~\bibnamefont {Ursin}}, \bibinfo {author} {\bibfnamefont {B.}~\bibnamefont {Wittmann}},\ and\ \bibinfo {author} {\bibfnamefont {A.}~\bibnamefont {Zeilinger}},\ }\bibfield  {title} {\bibinfo {title} {Significant-loophole-free test of bell's theorem with entangled photons},\ }\href {https://doi.org/10.1103/PhysRevLett.115.250401} {\bibfield  {journal} {\bibinfo  {journal} {Phys. Rev. Lett.}\ }\textbf {\bibinfo {volume} {115}},\ \bibinfo {pages} {250401} (\bibinfo {year} {2015})}\BibitemShut {NoStop}%
\bibitem [{\citenamefont {Shalm}\ \emph {et~al.}(2015)\citenamefont {Shalm}, \citenamefont {Meyer-Scott}, \citenamefont {Christensen}, \citenamefont {Bierhorst}, \citenamefont {Wayne}, \citenamefont {Stevens}, \citenamefont {Gerrits}, \citenamefont {Glancy}, \citenamefont {Hamel}, \citenamefont {Allman}, \citenamefont {Coakley}, \citenamefont {Dyer}, \citenamefont {Hodge}, \citenamefont {Lita}, \citenamefont {Verma}, \citenamefont {Lambrocco}, \citenamefont {Tortorici}, \citenamefont {Migdall}, \citenamefont {Zhang}, \citenamefont {Kumor}, \citenamefont {Farr}, \citenamefont {Marsili}, \citenamefont {Shaw}, \citenamefont {Stern}, \citenamefont {Abell\'an}, \citenamefont {Amaya}, \citenamefont {Pruneri}, \citenamefont {Jennewein}, \citenamefont {Mitchell}, \citenamefont {Kwiat}, \citenamefont {Bienfang}, \citenamefont {Mirin}, \citenamefont {Knill},\ and\ \citenamefont {Nam}}]{LoopholeFreeLocalRealism2015}%
  \BibitemOpen
  \bibfield  {author} {\bibinfo {author} {\bibfnamefont {L.~K.}\ \bibnamefont {Shalm}}, \bibinfo {author} {\bibfnamefont {E.}~\bibnamefont {Meyer-Scott}}, \bibinfo {author} {\bibfnamefont {B.~G.}\ \bibnamefont {Christensen}}, \bibinfo {author} {\bibfnamefont {P.}~\bibnamefont {Bierhorst}}, \bibinfo {author} {\bibfnamefont {M.~A.}\ \bibnamefont {Wayne}}, \bibinfo {author} {\bibfnamefont {M.~J.}\ \bibnamefont {Stevens}}, \bibinfo {author} {\bibfnamefont {T.}~\bibnamefont {Gerrits}}, \bibinfo {author} {\bibfnamefont {S.}~\bibnamefont {Glancy}}, \bibinfo {author} {\bibfnamefont {D.~R.}\ \bibnamefont {Hamel}}, \bibinfo {author} {\bibfnamefont {M.~S.}\ \bibnamefont {Allman}}, \bibinfo {author} {\bibfnamefont {K.~J.}\ \bibnamefont {Coakley}}, \bibinfo {author} {\bibfnamefont {S.~D.}\ \bibnamefont {Dyer}}, \bibinfo {author} {\bibfnamefont {C.}~\bibnamefont {Hodge}}, \bibinfo {author} {\bibfnamefont {A.~E.}\ \bibnamefont {Lita}}, \bibinfo {author} {\bibfnamefont {V.~B.}\ \bibnamefont {Verma}}, \bibinfo {author}
  {\bibfnamefont {C.}~\bibnamefont {Lambrocco}}, \bibinfo {author} {\bibfnamefont {E.}~\bibnamefont {Tortorici}}, \bibinfo {author} {\bibfnamefont {A.~L.}\ \bibnamefont {Migdall}}, \bibinfo {author} {\bibfnamefont {Y.}~\bibnamefont {Zhang}}, \bibinfo {author} {\bibfnamefont {D.~R.}\ \bibnamefont {Kumor}}, \bibinfo {author} {\bibfnamefont {W.~H.}\ \bibnamefont {Farr}}, \bibinfo {author} {\bibfnamefont {F.}~\bibnamefont {Marsili}}, \bibinfo {author} {\bibfnamefont {M.~D.}\ \bibnamefont {Shaw}}, \bibinfo {author} {\bibfnamefont {J.~A.}\ \bibnamefont {Stern}}, \bibinfo {author} {\bibfnamefont {C.}~\bibnamefont {Abell\'an}}, \bibinfo {author} {\bibfnamefont {W.}~\bibnamefont {Amaya}}, \bibinfo {author} {\bibfnamefont {V.}~\bibnamefont {Pruneri}}, \bibinfo {author} {\bibfnamefont {T.}~\bibnamefont {Jennewein}}, \bibinfo {author} {\bibfnamefont {M.~W.}\ \bibnamefont {Mitchell}}, \bibinfo {author} {\bibfnamefont {P.~G.}\ \bibnamefont {Kwiat}}, \bibinfo {author} {\bibfnamefont {J.~C.}\ \bibnamefont {Bienfang}},
  \bibinfo {author} {\bibfnamefont {R.~P.}\ \bibnamefont {Mirin}}, \bibinfo {author} {\bibfnamefont {E.}~\bibnamefont {Knill}},\ and\ \bibinfo {author} {\bibfnamefont {S.~W.}\ \bibnamefont {Nam}},\ }\bibfield  {title} {\bibinfo {title} {Strong loophole-free test of local realism},\ }\href {https://doi.org/10.1103/PhysRevLett.115.250402} {\bibfield  {journal} {\bibinfo  {journal} {Phys. Rev. Lett.}\ }\textbf {\bibinfo {volume} {115}},\ \bibinfo {pages} {250402} (\bibinfo {year} {2015})}\BibitemShut {NoStop}%
\bibitem [{\citenamefont {Rowe}\ \emph {et~al.}(2001)\citenamefont {Rowe}, \citenamefont {Kielpinski}, \citenamefont {Meyer}, \citenamefont {Sackett}, \citenamefont {Itano}, \citenamefont {Monroe},\ and\ \citenamefont {Wineland}}]{rowe2001ExperimentalViolationBells}%
  \BibitemOpen
  \bibfield  {author} {\bibinfo {author} {\bibfnamefont {M.~A.}\ \bibnamefont {Rowe}}, \bibinfo {author} {\bibfnamefont {D.}~\bibnamefont {Kielpinski}}, \bibinfo {author} {\bibfnamefont {V.}~\bibnamefont {Meyer}}, \bibinfo {author} {\bibfnamefont {C.~A.}\ \bibnamefont {Sackett}}, \bibinfo {author} {\bibfnamefont {W.~M.}\ \bibnamefont {Itano}}, \bibinfo {author} {\bibfnamefont {C.}~\bibnamefont {Monroe}},\ and\ \bibinfo {author} {\bibfnamefont {D.~J.}\ \bibnamefont {Wineland}},\ }\bibfield  {title} {\bibinfo {title} {Experimental violation of a {{Bell}}'s inequality with efficient detection},\ }\href {https://doi.org/10.1038/35057215} {\bibfield  {journal} {\bibinfo  {journal} {Nature}\ }\textbf {\bibinfo {volume} {409}},\ \bibinfo {pages} {791} (\bibinfo {year} {2001})}\BibitemShut {NoStop}%
\bibitem [{\citenamefont {Hensen}\ \emph {et~al.}(2015)\citenamefont {Hensen}, \citenamefont {Bernien}, \citenamefont {Dr{\'e}au}, \citenamefont {Reiserer}, \citenamefont {Kalb}, \citenamefont {Blok}, \citenamefont {Ruitenberg}, \citenamefont {Vermeulen}, \citenamefont {Schouten}, \citenamefont {Abell{\'a}n}, \citenamefont {Amaya}, \citenamefont {Pruneri}, \citenamefont {Mitchell}, \citenamefont {Markham}, \citenamefont {Twitchen}, \citenamefont {Elkouss}, \citenamefont {Wehner}, \citenamefont {Taminiau},\ and\ \citenamefont {Hanson}}]{hensen2015LoopholefreeBellInequality}%
  \BibitemOpen
  \bibfield  {author} {\bibinfo {author} {\bibfnamefont {B.}~\bibnamefont {Hensen}}, \bibinfo {author} {\bibfnamefont {H.}~\bibnamefont {Bernien}}, \bibinfo {author} {\bibfnamefont {A.~E.}\ \bibnamefont {Dr{\'e}au}}, \bibinfo {author} {\bibfnamefont {A.}~\bibnamefont {Reiserer}}, \bibinfo {author} {\bibfnamefont {N.}~\bibnamefont {Kalb}}, \bibinfo {author} {\bibfnamefont {M.~S.}\ \bibnamefont {Blok}}, \bibinfo {author} {\bibfnamefont {J.}~\bibnamefont {Ruitenberg}}, \bibinfo {author} {\bibfnamefont {R.~F.~L.}\ \bibnamefont {Vermeulen}}, \bibinfo {author} {\bibfnamefont {R.~N.}\ \bibnamefont {Schouten}}, \bibinfo {author} {\bibfnamefont {C.}~\bibnamefont {Abell{\'a}n}}, \bibinfo {author} {\bibfnamefont {W.}~\bibnamefont {Amaya}}, \bibinfo {author} {\bibfnamefont {V.}~\bibnamefont {Pruneri}}, \bibinfo {author} {\bibfnamefont {M.~W.}\ \bibnamefont {Mitchell}}, \bibinfo {author} {\bibfnamefont {M.}~\bibnamefont {Markham}}, \bibinfo {author} {\bibfnamefont {D.~J.}\ \bibnamefont {Twitchen}}, \bibinfo {author}
  {\bibfnamefont {D.}~\bibnamefont {Elkouss}}, \bibinfo {author} {\bibfnamefont {S.}~\bibnamefont {Wehner}}, \bibinfo {author} {\bibfnamefont {T.~H.}\ \bibnamefont {Taminiau}},\ and\ \bibinfo {author} {\bibfnamefont {R.}~\bibnamefont {Hanson}},\ }\bibfield  {title} {\bibinfo {title} {Loophole-free {{Bell}} inequality violation using electron spins separated by 1.3 kilometres},\ }\href {https://doi.org/10.1038/nature15759} {\bibfield  {journal} {\bibinfo  {journal} {Nature}\ }\textbf {\bibinfo {volume} {526}},\ \bibinfo {pages} {682} (\bibinfo {year} {2015})}\BibitemShut {NoStop}%
\bibitem [{\citenamefont {Schmied}\ \emph {et~al.}(2016)\citenamefont {Schmied}, \citenamefont {Bancal}, \citenamefont {Allard}, \citenamefont {Fadel}, \citenamefont {Scarani}, \citenamefont {Treutlein},\ and\ \citenamefont {Sangouard}}]{schmied2016BellCorrelationsBoseEinstein}%
  \BibitemOpen
  \bibfield  {author} {\bibinfo {author} {\bibfnamefont {R.}~\bibnamefont {Schmied}}, \bibinfo {author} {\bibfnamefont {J.-D.}\ \bibnamefont {Bancal}}, \bibinfo {author} {\bibfnamefont {B.}~\bibnamefont {Allard}}, \bibinfo {author} {\bibfnamefont {M.}~\bibnamefont {Fadel}}, \bibinfo {author} {\bibfnamefont {V.}~\bibnamefont {Scarani}}, \bibinfo {author} {\bibfnamefont {P.}~\bibnamefont {Treutlein}},\ and\ \bibinfo {author} {\bibfnamefont {N.}~\bibnamefont {Sangouard}},\ }\bibfield  {title} {\bibinfo {title} {Bell correlations in a {{Bose-Einstein}} condensate},\ }\href {https://doi.org/10.1126/science.aad8665} {\bibfield  {journal} {\bibinfo  {journal} {Science}\ }\textbf {\bibinfo {volume} {352}},\ \bibinfo {pages} {441} (\bibinfo {year} {2016})}\BibitemShut {NoStop}%
\bibitem [{\citenamefont {Pezz{\`e}}\ \emph {et~al.}(2018)\citenamefont {Pezz{\`e}}, \citenamefont {Smerzi}, \citenamefont {Oberthaler}, \citenamefont {Schmied},\ and\ \citenamefont {Treutlein}}]{pezze2018QuantumMetrologyNonclassical}%
  \BibitemOpen
  \bibfield  {author} {\bibinfo {author} {\bibfnamefont {L.}~\bibnamefont {Pezz{\`e}}}, \bibinfo {author} {\bibfnamefont {A.}~\bibnamefont {Smerzi}}, \bibinfo {author} {\bibfnamefont {M.~K.}\ \bibnamefont {Oberthaler}}, \bibinfo {author} {\bibfnamefont {R.}~\bibnamefont {Schmied}},\ and\ \bibinfo {author} {\bibfnamefont {P.}~\bibnamefont {Treutlein}},\ }\bibfield  {title} {\bibinfo {title} {Quantum metrology with nonclassical states of atomic ensembles},\ }\href {https://doi.org/10.1103/RevModPhys.90.035005} {\bibfield  {journal} {\bibinfo  {journal} {Rev. Mod. Phys.}\ }\textbf {\bibinfo {volume} {90}},\ \bibinfo {pages} {035005} (\bibinfo {year} {2018})}\BibitemShut {NoStop}%
\bibitem [{\citenamefont {Shin}\ \emph {et~al.}(2019)\citenamefont {Shin}, \citenamefont {Henson}, \citenamefont {Hodgman}, \citenamefont {Wasak}, \citenamefont {Chwede{\'n}czuk},\ and\ \citenamefont {Truscott}}]{shinBell2019d}%
  \BibitemOpen
  \bibfield  {author} {\bibinfo {author} {\bibfnamefont {D.~K.}\ \bibnamefont {Shin}}, \bibinfo {author} {\bibfnamefont {B.~M.}\ \bibnamefont {Henson}}, \bibinfo {author} {\bibfnamefont {S.~S.}\ \bibnamefont {Hodgman}}, \bibinfo {author} {\bibfnamefont {T.}~\bibnamefont {Wasak}}, \bibinfo {author} {\bibfnamefont {J.}~\bibnamefont {Chwede{\'n}czuk}},\ and\ \bibinfo {author} {\bibfnamefont {A.~G.}\ \bibnamefont {Truscott}},\ }\bibfield  {title} {\bibinfo {title} {Bell correlations between spatially separated pairs of atoms},\ }\href {https://doi.org/10.1038/s41467-019-12192-8} {\bibfield  {journal} {\bibinfo  {journal} {Nat. Commun.}\ }\textbf {\bibinfo {volume} {10}},\ \bibinfo {pages} {4447} (\bibinfo {year} {2019})}\BibitemShut {NoStop}%
\bibitem [{\citenamefont {Born}(1938)}]{Born1938suggestion}%
  \BibitemOpen
  \bibfield  {author} {\bibinfo {author} {\bibfnamefont {M.}~\bibnamefont {Born}},\ }\bibfield  {title} {\bibinfo {title} {A suggestion for unifying quantum theory and relativity},\ }\href {https://doi.org/10.1098/rspa.1938.0060} {\bibfield  {journal} {\bibinfo  {journal} {Proc. R. Soc. Lond. A}\ }\textbf {\bibinfo {volume} {165}},\ \bibinfo {pages} {291} (\bibinfo {year} {1938})}\BibitemShut {NoStop}%
\bibitem [{\citenamefont {Penrose}(1996)}]{penroseGravitys1996}%
  \BibitemOpen
  \bibfield  {author} {\bibinfo {author} {\bibfnamefont {R.}~\bibnamefont {Penrose}},\ }\bibfield  {title} {\bibinfo {title} {On {{Gravity}}'s role in {{Quantum State Reduction}}},\ }\href {https://doi.org/10.1007/BF02105068} {\bibfield  {journal} {\bibinfo  {journal} {Gen. Relat. Gravit.}\ }\textbf {\bibinfo {volume} {28}},\ \bibinfo {pages} {581} (\bibinfo {year} {1996})}\BibitemShut {NoStop}%
\bibitem [{\citenamefont {Khrennikov}(2017)}]{Khrennikov2017}%
  \BibitemOpen
  \bibfield  {author} {\bibinfo {author} {\bibfnamefont {A.}~\bibnamefont {Khrennikov}},\ }\bibfield  {title} {\bibinfo {title} {The present situation in quantum theory and its merging with general relativity},\ }\href {https://doi.org/10.1007/s10701-017-0089-0} {\bibfield  {journal} {\bibinfo  {journal} {Found. Phys.}\ }\textbf {\bibinfo {volume} {47}},\ \bibinfo {pages} {1077} (\bibinfo {year} {2017})}\BibitemShut {NoStop}%
\bibitem [{\citenamefont {Howl}\ \emph {et~al.}(2018)\citenamefont {Howl}, \citenamefont {Hackerm{\"u}ller}, \citenamefont {Bruschi},\ and\ \citenamefont {Fuentes}}]{Howl2018gravity}%
  \BibitemOpen
  \bibfield  {author} {\bibinfo {author} {\bibfnamefont {R.}~\bibnamefont {Howl}}, \bibinfo {author} {\bibfnamefont {L.}~\bibnamefont {Hackerm{\"u}ller}}, \bibinfo {author} {\bibfnamefont {D.~E.}\ \bibnamefont {Bruschi}},\ and\ \bibinfo {author} {\bibfnamefont {I.}~\bibnamefont {Fuentes}},\ }\bibfield  {title} {\bibinfo {title} {Gravity in the quantum lab},\ }\href {https://doi.org/10.1080/23746149.2017.1383184} {\bibfield  {journal} {\bibinfo  {journal} {Adv. Phys.}\ }\textbf {\bibinfo {volume} {3}},\ \bibinfo {pages} {1383184} (\bibinfo {year} {2018})}\BibitemShut {NoStop}%
\bibitem [{\citenamefont {Howl}\ \emph {et~al.}(2019)\citenamefont {Howl}, \citenamefont {Penrose},\ and\ \citenamefont {Fuentes}}]{Howl2019}%
  \BibitemOpen
  \bibfield  {author} {\bibinfo {author} {\bibfnamefont {R.}~\bibnamefont {Howl}}, \bibinfo {author} {\bibfnamefont {R.}~\bibnamefont {Penrose}},\ and\ \bibinfo {author} {\bibfnamefont {I.}~\bibnamefont {Fuentes}},\ }\bibfield  {title} {\bibinfo {title} {Exploring the unification of quantum theory and general relativity with a {B}ose–{E}instein condensate},\ }\href {https://doi.org/10.1088/1367-2630/ab104a} {\bibfield  {journal} {\bibinfo  {journal} {New J. Phys.}\ }\textbf {\bibinfo {volume} {21}},\ \bibinfo {pages} {043047} (\bibinfo {year} {2019})}\BibitemShut {NoStop}%
\bibitem [{\citenamefont {Ashtekar}\ and\ \citenamefont {Bianchi}(2021)}]{Ashtekar2021}%
  \BibitemOpen
  \bibfield  {author} {\bibinfo {author} {\bibfnamefont {A.}~\bibnamefont {Ashtekar}}\ and\ \bibinfo {author} {\bibfnamefont {E.}~\bibnamefont {Bianchi}},\ }\bibfield  {title} {\bibinfo {title} {A short review of loop quantum gravity},\ }\href {https://doi.org/10.1088/1361-6633/abed91} {\bibfield  {journal} {\bibinfo  {journal} {Rep. Prog. Phys.}\ }\textbf {\bibinfo {volume} {84}},\ \bibinfo {pages} {042001} (\bibinfo {year} {2021})}\BibitemShut {NoStop}%
\bibitem [{\citenamefont {Rarity}\ and\ \citenamefont {Tapster}(1990)}]{rarityExperimental1990}%
  \BibitemOpen
  \bibfield  {author} {\bibinfo {author} {\bibfnamefont {J.~G.}\ \bibnamefont {Rarity}}\ and\ \bibinfo {author} {\bibfnamefont {P.~R.}\ \bibnamefont {Tapster}},\ }\bibfield  {title} {\bibinfo {title} {Experimental violation of {{Bell}}'s inequality based on phase and momentum},\ }\href {https://doi.org/10.1103/PhysRevLett.64.2495} {\bibfield  {journal} {\bibinfo  {journal} {Phys. Rev. Lett.}\ }\textbf {\bibinfo {volume} {64}},\ \bibinfo {pages} {2495} (\bibinfo {year} {1990})}\BibitemShut {NoStop}%
\bibitem [{\citenamefont {Vassen}\ \emph {et~al.}(2012)\citenamefont {Vassen}, \citenamefont {{Cohen-Tannoudji}}, \citenamefont {Leduc}, \citenamefont {Boiron}, \citenamefont {Westbrook}, \citenamefont {Truscott}, \citenamefont {Baldwin}, \citenamefont {Birkl}, \citenamefont {Cancio},\ and\ \citenamefont {Trippenbach}}]{vassenCold2012}%
  \BibitemOpen
  \bibfield  {author} {\bibinfo {author} {\bibfnamefont {W.}~\bibnamefont {Vassen}}, \bibinfo {author} {\bibfnamefont {C.}~\bibnamefont {{Cohen-Tannoudji}}}, \bibinfo {author} {\bibfnamefont {M.}~\bibnamefont {Leduc}}, \bibinfo {author} {\bibfnamefont {D.}~\bibnamefont {Boiron}}, \bibinfo {author} {\bibfnamefont {C.~I.}\ \bibnamefont {Westbrook}}, \bibinfo {author} {\bibfnamefont {A.}~\bibnamefont {Truscott}}, \bibinfo {author} {\bibfnamefont {K.}~\bibnamefont {Baldwin}}, \bibinfo {author} {\bibfnamefont {G.}~\bibnamefont {Birkl}}, \bibinfo {author} {\bibfnamefont {P.}~\bibnamefont {Cancio}},\ and\ \bibinfo {author} {\bibfnamefont {M.}~\bibnamefont {Trippenbach}},\ }\bibfield  {title} {\bibinfo {title} {Cold and trapped metastable noble gases},\ }\href {https://doi.org/10.1103/RevModPhys.84.175} {\bibfield  {journal} {\bibinfo  {journal} {Rev. Mod. Phys.}\ }\textbf {\bibinfo {volume} {84}},\ \bibinfo {pages} {175} (\bibinfo {year} {2012})}\BibitemShut {NoStop}%
\bibitem [{\citenamefont {{Lewis-Swan}}\ and\ \citenamefont {Kheruntsyan}(2015)}]{lewis-swanProposal2015}%
  \BibitemOpen
  \bibfield  {author} {\bibinfo {author} {\bibfnamefont {R.~J.}\ \bibnamefont {{Lewis-Swan}}}\ and\ \bibinfo {author} {\bibfnamefont {K.~V.}\ \bibnamefont {Kheruntsyan}},\ }\bibfield  {title} {\bibinfo {title} {Proposal for a motional-state {{Bell}} inequality test with ultracold atoms},\ }\href {https://doi.org/10.1103/PhysRevA.91.052114} {\bibfield  {journal} {\bibinfo  {journal} {Phys. Rev. A}\ }\textbf {\bibinfo {volume} {91}},\ \bibinfo {pages} {052114} (\bibinfo {year} {2015})}\BibitemShut {NoStop}%
\bibitem [{\citenamefont {Kofler}\ \emph {et~al.}(2012)\citenamefont {Kofler}, \citenamefont {Singh}, \citenamefont {Ebner}, \citenamefont {Keller}, \citenamefont {Kotyrba},\ and\ \citenamefont {Zeilinger}}]{Kofler2012}%
  \BibitemOpen
  \bibfield  {author} {\bibinfo {author} {\bibfnamefont {J.}~\bibnamefont {Kofler}}, \bibinfo {author} {\bibfnamefont {M.}~\bibnamefont {Singh}}, \bibinfo {author} {\bibfnamefont {M.}~\bibnamefont {Ebner}}, \bibinfo {author} {\bibfnamefont {M.}~\bibnamefont {Keller}}, \bibinfo {author} {\bibfnamefont {M.}~\bibnamefont {Kotyrba}},\ and\ \bibinfo {author} {\bibfnamefont {A.}~\bibnamefont {Zeilinger}},\ }\bibfield  {title} {\bibinfo {title} {Einstein-podolsky-rosen correlations from colliding bose-einstein condensates},\ }\href {https://doi.org/10.1103/PhysRevA.86.032115} {\bibfield  {journal} {\bibinfo  {journal} {Phys. Rev. A}\ }\textbf {\bibinfo {volume} {86}},\ \bibinfo {pages} {032115} (\bibinfo {year} {2012})}\BibitemShut {NoStop}%
\bibitem [{\citenamefont {Keller}\ \emph {et~al.}(2014)\citenamefont {Keller}, \citenamefont {Kotyrba}, \citenamefont {Leupold}, \citenamefont {Singh}, \citenamefont {Ebner},\ and\ \citenamefont {Zeilinger}}]{Keller2014}%
  \BibitemOpen
  \bibfield  {author} {\bibinfo {author} {\bibfnamefont {M.}~\bibnamefont {Keller}}, \bibinfo {author} {\bibfnamefont {M.}~\bibnamefont {Kotyrba}}, \bibinfo {author} {\bibfnamefont {F.}~\bibnamefont {Leupold}}, \bibinfo {author} {\bibfnamefont {M.}~\bibnamefont {Singh}}, \bibinfo {author} {\bibfnamefont {M.}~\bibnamefont {Ebner}},\ and\ \bibinfo {author} {\bibfnamefont {A.}~\bibnamefont {Zeilinger}},\ }\bibfield  {title} {\bibinfo {title} {Bose-einstein condensate of metastable helium for quantum correlation experiments},\ }\href {https://doi.org/10.1103/PhysRevA.90.063607} {\bibfield  {journal} {\bibinfo  {journal} {Phys. Rev. A}\ }\textbf {\bibinfo {volume} {90}},\ \bibinfo {pages} {063607} (\bibinfo {year} {2014})}\BibitemShut {NoStop}%
\bibitem [{\citenamefont {Dussarrat}\ \emph {et~al.}(2017)\citenamefont {Dussarrat}, \citenamefont {Perrier}, \citenamefont {Imanaliev}, \citenamefont {Lopes}, \citenamefont {Aspect}, \citenamefont {Cheneau}, \citenamefont {Boiron},\ and\ \citenamefont {Westbrook}}]{dussarratTwoParticle2017}%
  \BibitemOpen
  \bibfield  {author} {\bibinfo {author} {\bibfnamefont {P.}~\bibnamefont {Dussarrat}}, \bibinfo {author} {\bibfnamefont {M.}~\bibnamefont {Perrier}}, \bibinfo {author} {\bibfnamefont {A.}~\bibnamefont {Imanaliev}}, \bibinfo {author} {\bibfnamefont {R.}~\bibnamefont {Lopes}}, \bibinfo {author} {\bibfnamefont {A.}~\bibnamefont {Aspect}}, \bibinfo {author} {\bibfnamefont {M.}~\bibnamefont {Cheneau}}, \bibinfo {author} {\bibfnamefont {D.}~\bibnamefont {Boiron}},\ and\ \bibinfo {author} {\bibfnamefont {C.~I.}\ \bibnamefont {Westbrook}},\ }\bibfield  {title} {\bibinfo {title} {Two-{{Particle Four-Mode Interferometer}} for {{Atoms}}},\ }\href {https://doi.org/10.1103/PhysRevLett.119.173202} {\bibfield  {journal} {\bibinfo  {journal} {Phys. Rev. Lett.}\ }\textbf {\bibinfo {volume} {119}},\ \bibinfo {pages} {173202} (\bibinfo {year} {2017})}\BibitemShut {NoStop}%
\bibitem [{\citenamefont {Thomas}\ \emph {et~al.}(2022)\citenamefont {Thomas}, \citenamefont {Henson}, \citenamefont {Wang}, \citenamefont {{Lewis-Swan}}, \citenamefont {Kheruntsyan}, \citenamefont {Hodgman},\ and\ \citenamefont {Truscott}}]{thomasMatterwave2022}%
  \BibitemOpen
  \bibfield  {author} {\bibinfo {author} {\bibfnamefont {K.~F.}\ \bibnamefont {Thomas}}, \bibinfo {author} {\bibfnamefont {B.~M.}\ \bibnamefont {Henson}}, \bibinfo {author} {\bibfnamefont {Y.}~\bibnamefont {Wang}}, \bibinfo {author} {\bibfnamefont {R.~J.}\ \bibnamefont {{Lewis-Swan}}}, \bibinfo {author} {\bibfnamefont {K.~V.}\ \bibnamefont {Kheruntsyan}}, \bibinfo {author} {\bibfnamefont {S.~S.}\ \bibnamefont {Hodgman}},\ and\ \bibinfo {author} {\bibfnamefont {A.~G.}\ \bibnamefont {Truscott}},\ }\bibfield  {title} {\bibinfo {title} {A matter-wave {{Rarity}}--{{Tapster}} interferometer to demonstrate non-locality},\ }\href {https://doi.org/10.1140/epjd/s10053-022-00551-y} {\bibfield  {journal} {\bibinfo  {journal} {Eur. Phys. J. D}\ }\textbf {\bibinfo {volume} {76}},\ \bibinfo {pages} {244} (\bibinfo {year} {2022})}\BibitemShut {NoStop}%
\bibitem [{\citenamefont {Leprince}\ \emph {et~al.}(2024)\citenamefont {Leprince}, \citenamefont {Gondret}, \citenamefont {Lamirault}, \citenamefont {Dias}, \citenamefont {Marolleau}, \citenamefont {Boiron},\ and\ \citenamefont {Westbrook}}]{Leprince2024}%
  \BibitemOpen
  \bibfield  {author} {\bibinfo {author} {\bibfnamefont {C.}~\bibnamefont {Leprince}}, \bibinfo {author} {\bibfnamefont {V.}~\bibnamefont {Gondret}}, \bibinfo {author} {\bibfnamefont {C.}~\bibnamefont {Lamirault}}, \bibinfo {author} {\bibfnamefont {R.}~\bibnamefont {Dias}}, \bibinfo {author} {\bibfnamefont {Q.}~\bibnamefont {Marolleau}}, \bibinfo {author} {\bibfnamefont {D.}~\bibnamefont {Boiron}},\ and\ \bibinfo {author} {\bibfnamefont {C.~I.}\ \bibnamefont {Westbrook}},\ }\bibfield  {title} {\bibinfo {title} {Coherent coupling of momentum states: selectivity and phase control},\ }\href {https://arxiv.org/abs/2411.09284} {\bibfield  {journal} {\bibinfo  {journal} {arXiv:2411.09284}\ } (\bibinfo {year} {2024})}\BibitemShut {NoStop}%
\bibitem [{\citenamefont {Perrier}\ \emph {et~al.}(2019)\citenamefont {Perrier}, \citenamefont {Amodjee}, \citenamefont {Dussarrat}, \citenamefont {Dareau}, \citenamefont {Aspect}, \citenamefont {Cheneau}, \citenamefont {Boiron},\ and\ \citenamefont {Westbrook}}]{Perrier2019}%
  \BibitemOpen
  \bibfield  {author} {\bibinfo {author} {\bibfnamefont {M.}~\bibnamefont {Perrier}}, \bibinfo {author} {\bibfnamefont {Z.}~\bibnamefont {Amodjee}}, \bibinfo {author} {\bibfnamefont {P.}~\bibnamefont {Dussarrat}}, \bibinfo {author} {\bibfnamefont {A.}~\bibnamefont {Dareau}}, \bibinfo {author} {\bibfnamefont {A.}~\bibnamefont {Aspect}}, \bibinfo {author} {\bibfnamefont {M.}~\bibnamefont {Cheneau}}, \bibinfo {author} {\bibfnamefont {D.}~\bibnamefont {Boiron}},\ and\ \bibinfo {author} {\bibfnamefont {C.~I.}\ \bibnamefont {Westbrook}},\ }\bibfield  {title} {\bibinfo {title} {{Thermal counting statistics in an atomic two-mode squeezed vacuum state}},\ }\href {https://doi.org/10.21468/SciPostPhys.7.1.002} {\bibfield  {journal} {\bibinfo  {journal} {SciPost Phys.}\ }\textbf {\bibinfo {volume} {7}},\ \bibinfo {pages} {002} (\bibinfo {year} {2019})}\BibitemShut {NoStop}%
\bibitem [{\citenamefont {Perrin}\ \emph {et~al.}(2007)\citenamefont {Perrin}, \citenamefont {Chang}, \citenamefont {Krachmalnicoff}, \citenamefont {Schellekens}, \citenamefont {Boiron}, \citenamefont {Aspect},\ and\ \citenamefont {Westbrook}}]{perrinObservation2007}%
  \BibitemOpen
  \bibfield  {author} {\bibinfo {author} {\bibfnamefont {A.}~\bibnamefont {Perrin}}, \bibinfo {author} {\bibfnamefont {H.}~\bibnamefont {Chang}}, \bibinfo {author} {\bibfnamefont {V.}~\bibnamefont {Krachmalnicoff}}, \bibinfo {author} {\bibfnamefont {M.}~\bibnamefont {Schellekens}}, \bibinfo {author} {\bibfnamefont {D.}~\bibnamefont {Boiron}}, \bibinfo {author} {\bibfnamefont {A.}~\bibnamefont {Aspect}},\ and\ \bibinfo {author} {\bibfnamefont {C.~I.}\ \bibnamefont {Westbrook}},\ }\bibfield  {title} {\bibinfo {title} {Observation of {{Atom Pairs}} in {{Spontaneous Four-Wave Mixing}} of {{Two Colliding Bose-Einstein Condensates}}},\ }\href {https://doi.org/10.1103/PhysRevLett.99.150405} {\bibfield  {journal} {\bibinfo  {journal} {Phys. Rev. Lett.}\ }\textbf {\bibinfo {volume} {99}},\ \bibinfo {pages} {150405} (\bibinfo {year} {2007})}\BibitemShut {NoStop}%
\bibitem [{\citenamefont {Perrin}\ \emph {et~al.}(2008)\citenamefont {Perrin}, \citenamefont {Savage}, \citenamefont {Boiron}, \citenamefont {Krachmalnicoff}, \citenamefont {Westbrook},\ and\ \citenamefont {Kheruntsyan}}]{perrinAtomic2008}%
  \BibitemOpen
  \bibfield  {author} {\bibinfo {author} {\bibfnamefont {A.}~\bibnamefont {Perrin}}, \bibinfo {author} {\bibfnamefont {C.~M.}\ \bibnamefont {Savage}}, \bibinfo {author} {\bibfnamefont {D.}~\bibnamefont {Boiron}}, \bibinfo {author} {\bibfnamefont {V.}~\bibnamefont {Krachmalnicoff}}, \bibinfo {author} {\bibfnamefont {C.~I.}\ \bibnamefont {Westbrook}},\ and\ \bibinfo {author} {\bibfnamefont {K.}~\bibnamefont {Kheruntsyan}},\ }\bibfield  {title} {\bibinfo {title} {Atomic four-wave mixing via condensate collisions},\ }\href {https://doi.org/10.1088/1367-2630/10/4/045021} {\bibfield  {journal} {\bibinfo  {journal} {New J. Phys.}\ }\textbf {\bibinfo {volume} {10}},\ \bibinfo {pages} {045021} (\bibinfo {year} {2008})}\BibitemShut {NoStop}%
\bibitem [{\citenamefont {Khakimov}\ \emph {et~al.}(2016)\citenamefont {Khakimov}, \citenamefont {Henson}, \citenamefont {Shin}, \citenamefont {Hodgman}, \citenamefont {Dall}, \citenamefont {Baldwin},\ and\ \citenamefont {Truscott}}]{khakimovGhost2016a}%
  \BibitemOpen
  \bibfield  {author} {\bibinfo {author} {\bibfnamefont {R.~I.}\ \bibnamefont {Khakimov}}, \bibinfo {author} {\bibfnamefont {B.~M.}\ \bibnamefont {Henson}}, \bibinfo {author} {\bibfnamefont {D.~K.}\ \bibnamefont {Shin}}, \bibinfo {author} {\bibfnamefont {S.~S.}\ \bibnamefont {Hodgman}}, \bibinfo {author} {\bibfnamefont {R.~G.}\ \bibnamefont {Dall}}, \bibinfo {author} {\bibfnamefont {K.~G.~H.}\ \bibnamefont {Baldwin}},\ and\ \bibinfo {author} {\bibfnamefont {A.~G.}\ \bibnamefont {Truscott}},\ }\bibfield  {title} {\bibinfo {title} {Ghost imaging with atoms},\ }\href {https://doi.org/10.1038/nature20154} {\bibfield  {journal} {\bibinfo  {journal} {Nature}\ }\textbf {\bibinfo {volume} {540}},\ \bibinfo {pages} {100} (\bibinfo {year} {2016})}\BibitemShut {NoStop}%
\bibitem [{\citenamefont {Hilligs\o{}e}\ and\ \citenamefont {M\o{}lmer}(2005)}]{Hilligsoe2005}%
  \BibitemOpen
  \bibfield  {author} {\bibinfo {author} {\bibfnamefont {K.~M.}\ \bibnamefont {Hilligs\o{}e}}\ and\ \bibinfo {author} {\bibfnamefont {K.}~\bibnamefont {M\o{}lmer}},\ }\bibfield  {title} {\bibinfo {title} {Phase-matched four wave mixing and quantum beam splitting of matter waves in a periodic potential},\ }\href {https://doi.org/10.1103/PhysRevA.71.041602} {\bibfield  {journal} {\bibinfo  {journal} {Phys. Rev. A}\ }\textbf {\bibinfo {volume} {71}},\ \bibinfo {pages} {041602} (\bibinfo {year} {2005})}\BibitemShut {NoStop}%
\bibitem [{\citenamefont {Campbell}\ \emph {et~al.}(2006)\citenamefont {Campbell}, \citenamefont {Mun}, \citenamefont {Boyd}, \citenamefont {Streed}, \citenamefont {Ketterle},\ and\ \citenamefont {Pritchard}}]{Campbell2006}%
  \BibitemOpen
  \bibfield  {author} {\bibinfo {author} {\bibfnamefont {G.~K.}\ \bibnamefont {Campbell}}, \bibinfo {author} {\bibfnamefont {J.}~\bibnamefont {Mun}}, \bibinfo {author} {\bibfnamefont {M.}~\bibnamefont {Boyd}}, \bibinfo {author} {\bibfnamefont {E.~W.}\ \bibnamefont {Streed}}, \bibinfo {author} {\bibfnamefont {W.}~\bibnamefont {Ketterle}},\ and\ \bibinfo {author} {\bibfnamefont {D.~E.}\ \bibnamefont {Pritchard}},\ }\bibfield  {title} {\bibinfo {title} {Parametric amplification of scattered atom pairs},\ }\href {https://doi.org/10.1103/PhysRevLett.96.020406} {\bibfield  {journal} {\bibinfo  {journal} {Phys. Rev. Lett.}\ }\textbf {\bibinfo {volume} {96}},\ \bibinfo {pages} {020406} (\bibinfo {year} {2006})}\BibitemShut {NoStop}%
\bibitem [{\citenamefont {B{\"u}cker}\ \emph {et~al.}(2011)\citenamefont {B{\"u}cker}, \citenamefont {Grond}, \citenamefont {Manz}, \citenamefont {Berrada}, \citenamefont {Betz}, \citenamefont {Koller}, \citenamefont {Hohenester}, \citenamefont {Schumm}, \citenamefont {Perrin},\ and\ \citenamefont {Schmiedmayer}}]{Bucker2011}%
  \BibitemOpen
  \bibfield  {author} {\bibinfo {author} {\bibfnamefont {R.}~\bibnamefont {B{\"u}cker}}, \bibinfo {author} {\bibfnamefont {J.}~\bibnamefont {Grond}}, \bibinfo {author} {\bibfnamefont {S.}~\bibnamefont {Manz}}, \bibinfo {author} {\bibfnamefont {T.}~\bibnamefont {Berrada}}, \bibinfo {author} {\bibfnamefont {T.}~\bibnamefont {Betz}}, \bibinfo {author} {\bibfnamefont {C.}~\bibnamefont {Koller}}, \bibinfo {author} {\bibfnamefont {U.}~\bibnamefont {Hohenester}}, \bibinfo {author} {\bibfnamefont {T.}~\bibnamefont {Schumm}}, \bibinfo {author} {\bibfnamefont {A.}~\bibnamefont {Perrin}},\ and\ \bibinfo {author} {\bibfnamefont {J.}~\bibnamefont {Schmiedmayer}},\ }\bibfield  {title} {\bibinfo {title} {Twin-atom beams},\ }\href {https://doi.org/10.1038/nphys1992} {\bibfield  {journal} {\bibinfo  {journal} {Nat. Phys.}\ }\textbf {\bibinfo {volume} {7}},\ \bibinfo {pages} {608} (\bibinfo {year} {2011})}\BibitemShut {NoStop}%
\bibitem [{\citenamefont {Bonneau}\ \emph {et~al.}(2013)\citenamefont {Bonneau}, \citenamefont {Ruaudel}, \citenamefont {Lopes}, \citenamefont {Jaskula}, \citenamefont {Aspect}, \citenamefont {Boiron},\ and\ \citenamefont {Westbrook}}]{Bonneau2013}%
  \BibitemOpen
  \bibfield  {author} {\bibinfo {author} {\bibfnamefont {M.}~\bibnamefont {Bonneau}}, \bibinfo {author} {\bibfnamefont {J.}~\bibnamefont {Ruaudel}}, \bibinfo {author} {\bibfnamefont {R.}~\bibnamefont {Lopes}}, \bibinfo {author} {\bibfnamefont {J.-C.}\ \bibnamefont {Jaskula}}, \bibinfo {author} {\bibfnamefont {A.}~\bibnamefont {Aspect}}, \bibinfo {author} {\bibfnamefont {D.}~\bibnamefont {Boiron}},\ and\ \bibinfo {author} {\bibfnamefont {C.~I.}\ \bibnamefont {Westbrook}},\ }\bibfield  {title} {\bibinfo {title} {Tunable source of correlated atom beams},\ }\href {https://doi.org/10.1103/PhysRevA.87.061603} {\bibfield  {journal} {\bibinfo  {journal} {Phys. Rev. A}\ }\textbf {\bibinfo {volume} {87}},\ \bibinfo {pages} {061603} (\bibinfo {year} {2013})}\BibitemShut {NoStop}%
\bibitem [{\citenamefont {Lopes}\ \emph {et~al.}(2015)\citenamefont {Lopes}, \citenamefont {Imanaliev}, \citenamefont {Aspect}, \citenamefont {Cheneau}, \citenamefont {Boiron},\ and\ \citenamefont {Westbrook}}]{Lopes2015}%
  \BibitemOpen
  \bibfield  {author} {\bibinfo {author} {\bibfnamefont {R.}~\bibnamefont {Lopes}}, \bibinfo {author} {\bibfnamefont {A.}~\bibnamefont {Imanaliev}}, \bibinfo {author} {\bibfnamefont {A.}~\bibnamefont {Aspect}}, \bibinfo {author} {\bibfnamefont {M.}~\bibnamefont {Cheneau}}, \bibinfo {author} {\bibfnamefont {D.}~\bibnamefont {Boiron}},\ and\ \bibinfo {author} {\bibfnamefont {C.~I.}\ \bibnamefont {Westbrook}},\ }\bibfield  {title} {\bibinfo {title} {Atomic hong--ou--mandel experiment},\ }\href {https://doi.org/10.1038/nature14331} {\bibfield  {journal} {\bibinfo  {journal} {Nature}\ }\textbf {\bibinfo {volume} {520}},\ \bibinfo {pages} {66} (\bibinfo {year} {2015})}\BibitemShut {NoStop}%
\bibitem [{\citenamefont {Hodgman}\ \emph {et~al.}(2009)\citenamefont {Hodgman}, \citenamefont {Dall}, \citenamefont {Byron}, \citenamefont {Baldwin}, \citenamefont {Buckman},\ and\ \citenamefont {Truscott}}]{hodgmanMetastable2009}%
  \BibitemOpen
  \bibfield  {author} {\bibinfo {author} {\bibfnamefont {S.~S.}\ \bibnamefont {Hodgman}}, \bibinfo {author} {\bibfnamefont {R.~G.}\ \bibnamefont {Dall}}, \bibinfo {author} {\bibfnamefont {L.~J.}\ \bibnamefont {Byron}}, \bibinfo {author} {\bibfnamefont {K.~G.~H.}\ \bibnamefont {Baldwin}}, \bibinfo {author} {\bibfnamefont {S.~J.}\ \bibnamefont {Buckman}},\ and\ \bibinfo {author} {\bibfnamefont {A.~G.}\ \bibnamefont {Truscott}},\ }\bibfield  {title} {\bibinfo {title} {Metastable {{Helium}}: {{A New Determination}} of the {{Longest Atomic Excited-State Lifetime}}},\ }\href {https://doi.org/10.1103/PhysRevLett.103.053002} {\bibfield  {journal} {\bibinfo  {journal} {Phys. Rev. Lett.}\ }\textbf {\bibinfo {volume} {103}},\ \bibinfo {pages} {053002} (\bibinfo {year} {2009})}\BibitemShut {NoStop}%
\bibitem [{\citenamefont {Dall}\ and\ \citenamefont {Truscott}(2007)}]{dallBose2007a}%
  \BibitemOpen
  \bibfield  {author} {\bibinfo {author} {\bibfnamefont {R.~G.}\ \bibnamefont {Dall}}\ and\ \bibinfo {author} {\bibfnamefont {A.~G.}\ \bibnamefont {Truscott}},\ }\bibfield  {title} {\bibinfo {title} {Bose--{{Einstein}} condensation of metastable helium in a bi-planar quadrupole {{Ioffe}} configuration trap},\ }\href {https://doi.org/10.1016/j.optcom.2006.09.031} {\bibfield  {journal} {\bibinfo  {journal} {Opt. Commun.}\ }\textbf {\bibinfo {volume} {270}},\ \bibinfo {pages} {255} (\bibinfo {year} {2007})}\BibitemShut {NoStop}%
\bibitem [{\citenamefont {Moler}\ \emph {et~al.}(1992)\citenamefont {Moler}, \citenamefont {Weiss}, \citenamefont {Kasevich},\ and\ \citenamefont {Chu}}]{molerTheoretical1992}%
  \BibitemOpen
  \bibfield  {author} {\bibinfo {author} {\bibfnamefont {K.}~\bibnamefont {Moler}}, \bibinfo {author} {\bibfnamefont {D.~S.}\ \bibnamefont {Weiss}}, \bibinfo {author} {\bibfnamefont {M.}~\bibnamefont {Kasevich}},\ and\ \bibinfo {author} {\bibfnamefont {S.}~\bibnamefont {Chu}},\ }\bibfield  {title} {\bibinfo {title} {Theoretical analysis of velocity-selective {{Raman}} transitions},\ }\href {https://doi.org/10.1103/PhysRevA.45.342} {\bibfield  {journal} {\bibinfo  {journal} {Phys. Rev. A}\ }\textbf {\bibinfo {volume} {45}},\ \bibinfo {pages} {342} (\bibinfo {year} {1992})}\BibitemShut {NoStop}%
\bibitem [{\citenamefont {Kozuma}\ \emph {et~al.}(1999)\citenamefont {Kozuma}, \citenamefont {Deng}, \citenamefont {Hagley}, \citenamefont {Wen}, \citenamefont {Lutwak}, \citenamefont {Helmerson}, \citenamefont {Rolston},\ and\ \citenamefont {Phillips}}]{kozumaCoherent1999}%
  \BibitemOpen
  \bibfield  {author} {\bibinfo {author} {\bibfnamefont {M.}~\bibnamefont {Kozuma}}, \bibinfo {author} {\bibfnamefont {L.}~\bibnamefont {Deng}}, \bibinfo {author} {\bibfnamefont {E.~W.}\ \bibnamefont {Hagley}}, \bibinfo {author} {\bibfnamefont {J.}~\bibnamefont {Wen}}, \bibinfo {author} {\bibfnamefont {R.}~\bibnamefont {Lutwak}}, \bibinfo {author} {\bibfnamefont {K.}~\bibnamefont {Helmerson}}, \bibinfo {author} {\bibfnamefont {S.~L.}\ \bibnamefont {Rolston}},\ and\ \bibinfo {author} {\bibfnamefont {W.~D.}\ \bibnamefont {Phillips}},\ }\bibfield  {title} {\bibinfo {title} {Coherent {{Splitting}} of {{Bose-Einstein Condensed Atoms}} with {{Optically Induced Bragg Diffraction}}},\ }\href {https://doi.org/10.1103/PhysRevLett.82.871} {\bibfield  {journal} {\bibinfo  {journal} {Phys. Rev. Lett.}\ }\textbf {\bibinfo {volume} {82}},\ \bibinfo {pages} {871} (\bibinfo {year} {1999})}\BibitemShut {NoStop}%
\bibitem [{\citenamefont {Gupta}\ \emph {et~al.}(2001)\citenamefont {Gupta}, \citenamefont {Leanhardt}, \citenamefont {Cronin},\ and\ \citenamefont {Pritchard}}]{guptaCoherent2001}%
  \BibitemOpen
  \bibfield  {author} {\bibinfo {author} {\bibfnamefont {S.}~\bibnamefont {Gupta}}, \bibinfo {author} {\bibfnamefont {A.~E.}\ \bibnamefont {Leanhardt}}, \bibinfo {author} {\bibfnamefont {A.~D.}\ \bibnamefont {Cronin}},\ and\ \bibinfo {author} {\bibfnamefont {D.~E.}\ \bibnamefont {Pritchard}},\ }\bibfield  {title} {\bibinfo {title} {Coherent manipulation of atoms with standing light waves},\ }\href@noop {} {\bibfield  {journal} {\bibinfo  {journal} {Comptes Rendus de l'Acad{\'e}mie des Sciences - Series IV - Physics}\ }\textbf {\bibinfo {volume} {2}},\ \bibinfo {pages} {479} (\bibinfo {year} {2001})}\BibitemShut {NoStop}%
\bibitem [{\citenamefont {RuGway}\ \emph {et~al.}(2011)\citenamefont {RuGway}, \citenamefont {Hodgman}, \citenamefont {Dall}, \citenamefont {Johnsson},\ and\ \citenamefont {Truscott}}]{rugwayCorrelations2011a}%
  \BibitemOpen
  \bibfield  {author} {\bibinfo {author} {\bibfnamefont {W.}~\bibnamefont {RuGway}}, \bibinfo {author} {\bibfnamefont {S.~S.}\ \bibnamefont {Hodgman}}, \bibinfo {author} {\bibfnamefont {R.~G.}\ \bibnamefont {Dall}}, \bibinfo {author} {\bibfnamefont {M.~T.}\ \bibnamefont {Johnsson}},\ and\ \bibinfo {author} {\bibfnamefont {A.~G.}\ \bibnamefont {Truscott}},\ }\bibfield  {title} {\bibinfo {title} {Correlations in {{Amplified Four-Wave Mixing}} of {{Matter Waves}}},\ }\href {https://doi.org/10.1103/PhysRevLett.107.075301} {\bibfield  {journal} {\bibinfo  {journal} {Phys. Rev. Lett.}\ }\textbf {\bibinfo {volume} {107}},\ \bibinfo {pages} {075301} (\bibinfo {year} {2011})}\BibitemShut {NoStop}%
\bibitem [{\citenamefont {Couteau}(2018)}]{couteauSpontaneous2018}%
  \BibitemOpen
  \bibfield  {author} {\bibinfo {author} {\bibfnamefont {C.}~\bibnamefont {Couteau}},\ }\bibfield  {title} {\bibinfo {title} {Spontaneous parametric down-conversion},\ }\href {https://doi.org/10.1080/00107514.2018.1488463} {\bibfield  {journal} {\bibinfo  {journal} {Contemp. Phys.}\ }\textbf {\bibinfo {volume} {59}},\ \bibinfo {pages} {291} (\bibinfo {year} {2018})}\BibitemShut {NoStop}%
\bibitem [{\citenamefont {Hodgman}\ \emph {et~al.}(2017)\citenamefont {Hodgman}, \citenamefont {Khakimov}, \citenamefont {{Lewis-Swan}}, \citenamefont {Truscott},\ and\ \citenamefont {Kheruntsyan}}]{hodgmanSolving2017}%
  \BibitemOpen
  \bibfield  {author} {\bibinfo {author} {\bibfnamefont {S.~S.}\ \bibnamefont {Hodgman}}, \bibinfo {author} {\bibfnamefont {R.~I.}\ \bibnamefont {Khakimov}}, \bibinfo {author} {\bibfnamefont {R.~J.}\ \bibnamefont {{Lewis-Swan}}}, \bibinfo {author} {\bibfnamefont {A.~G.}\ \bibnamefont {Truscott}},\ and\ \bibinfo {author} {\bibfnamefont {K.~V.}\ \bibnamefont {Kheruntsyan}},\ }\bibfield  {title} {\bibinfo {title} {Solving the {{Quantum Many-Body Problem}} via {{Correlations Measured}} with a {{Momentum Microscope}}},\ }\href {https://doi.org/10.1103/PhysRevLett.118.240402} {\bibfield  {journal} {\bibinfo  {journal} {Phys. Rev. Lett.}\ }\textbf {\bibinfo {volume} {118}},\ \bibinfo {pages} {240402} (\bibinfo {year} {2017})}\BibitemShut {NoStop}%
\bibitem [{SOM()}]{SOM}%
  \BibitemOpen
  \href@noop {} {\bibinfo {title} {See supplementary material for more information}}\BibitemShut {NoStop}%
\bibitem [{\citenamefont {Adams}\ \emph {et~al.}(1994)\citenamefont {Adams}, \citenamefont {Sigel},\ and\ \citenamefont {Mlynek}}]{adamsAtom1994}%
  \BibitemOpen
  \bibfield  {author} {\bibinfo {author} {\bibfnamefont {C.~S.}\ \bibnamefont {Adams}}, \bibinfo {author} {\bibfnamefont {M.}~\bibnamefont {Sigel}},\ and\ \bibinfo {author} {\bibfnamefont {J.}~\bibnamefont {Mlynek}},\ }\bibfield  {title} {\bibinfo {title} {Atom optics},\ }\href {https://doi.org/10.1016/0370-1573(94)90066-3} {\bibfield  {journal} {\bibinfo  {journal} {Phys. Rep.}\ }\textbf {\bibinfo {volume} {240}},\ \bibinfo {pages} {143} (\bibinfo {year} {1994})}\BibitemShut {NoStop}%
\bibitem [{\citenamefont {Manning}\ \emph {et~al.}(2010)\citenamefont {Manning}, \citenamefont {Hodgman}, \citenamefont {Dall}, \citenamefont {Johnsson},\ and\ \citenamefont {Truscott}}]{manningHanbury2010}%
  \BibitemOpen
  \bibfield  {author} {\bibinfo {author} {\bibfnamefont {A.~G.}\ \bibnamefont {Manning}}, \bibinfo {author} {\bibfnamefont {S.~S.}\ \bibnamefont {Hodgman}}, \bibinfo {author} {\bibfnamefont {R.~G.}\ \bibnamefont {Dall}}, \bibinfo {author} {\bibfnamefont {M.~T.}\ \bibnamefont {Johnsson}},\ and\ \bibinfo {author} {\bibfnamefont {A.~G.}\ \bibnamefont {Truscott}},\ }\bibfield  {title} {\bibinfo {title} {The {{Hanbury Brown-Twiss}} effect in a pulsed atom laser},\ }\href {https://doi.org/10.1364/OE.18.018712} {\bibfield  {journal} {\bibinfo  {journal} {Opt. Express}\ }\textbf {\bibinfo {volume} {18}},\ \bibinfo {pages} {18712} (\bibinfo {year} {2010})}\BibitemShut {NoStop}%
\bibitem [{\citenamefont {Glauber}(1963)}]{glauberQuantum1963b}%
  \BibitemOpen
  \bibfield  {author} {\bibinfo {author} {\bibfnamefont {R.~J.}\ \bibnamefont {Glauber}},\ }\bibfield  {title} {\bibinfo {title} {The {{Quantum Theory}} of {{Optical Coherence}}},\ }\href {https://doi.org/10.1103/PhysRev.130.2529} {\bibfield  {journal} {\bibinfo  {journal} {Phys. Rev.}\ }\textbf {\bibinfo {volume} {130}},\ \bibinfo {pages} {2529} (\bibinfo {year} {1963})}\BibitemShut {NoStop}%
\bibitem [{\citenamefont {{\"O}gren}\ and\ \citenamefont {Kheruntsyan}(2009)}]{ogrenAtomatom2009}%
  \BibitemOpen
  \bibfield  {author} {\bibinfo {author} {\bibfnamefont {M.}~\bibnamefont {{\"O}gren}}\ and\ \bibinfo {author} {\bibfnamefont {K.~V.}\ \bibnamefont {Kheruntsyan}},\ }\bibfield  {title} {\bibinfo {title} {Atom-atom correlations in colliding {{Bose-Einstein}} condensates},\ }\href {https://doi.org/10.1103/PhysRevA.79.021606} {\bibfield  {journal} {\bibinfo  {journal} {Phys. Rev. A}\ }\textbf {\bibinfo {volume} {79}},\ \bibinfo {pages} {021606} (\bibinfo {year} {2009})}\BibitemShut {NoStop}%
\bibitem [{\citenamefont {Wasak}\ and\ \citenamefont {Chwede{\'n}czuk}(2018)}]{wasakBell2018}%
  \BibitemOpen
  \bibfield  {author} {\bibinfo {author} {\bibfnamefont {T.}~\bibnamefont {Wasak}}\ and\ \bibinfo {author} {\bibfnamefont {J.}~\bibnamefont {Chwede{\'n}czuk}},\ }\bibfield  {title} {\bibinfo {title} {Bell {{Inequality}}, {{Einstein-Podolsky-Rosen Steering}}, and {{Quantum Metrology}} with {{Spinor Bose-Einstein Condensates}}},\ }\href {https://doi.org/10.1103/PhysRevLett.120.140406} {\bibfield  {journal} {\bibinfo  {journal} {Phys. Rev. Lett.}\ }\textbf {\bibinfo {volume} {120}},\ \bibinfo {pages} {140406} (\bibinfo {year} {2018})}\BibitemShut {NoStop}%
\bibitem [{\citenamefont {Clauser}\ \emph {et~al.}(1969)\citenamefont {Clauser}, \citenamefont {Horne}, \citenamefont {Shimony},\ and\ \citenamefont {Holt}}]{clauserProposed1969}%
  \BibitemOpen
  \bibfield  {author} {\bibinfo {author} {\bibfnamefont {J.~F.}\ \bibnamefont {Clauser}}, \bibinfo {author} {\bibfnamefont {M.~A.}\ \bibnamefont {Horne}}, \bibinfo {author} {\bibfnamefont {A.}~\bibnamefont {Shimony}},\ and\ \bibinfo {author} {\bibfnamefont {R.~A.}\ \bibnamefont {Holt}},\ }\bibfield  {title} {\bibinfo {title} {Proposed {{Experiment}} to {{Test Local Hidden-Variable Theories}}},\ }\href {https://doi.org/10.1103/PhysRevLett.23.880} {\bibfield  {journal} {\bibinfo  {journal} {Phys. Rev. Lett.}\ }\textbf {\bibinfo {volume} {23}},\ \bibinfo {pages} {880} (\bibinfo {year} {1969})}\BibitemShut {NoStop}%
\bibitem [{\citenamefont {Clauser}\ and\ \citenamefont {Shimony}(1978)}]{Clauser_Shimony_review_1978}%
  \BibitemOpen
  \bibfield  {author} {\bibinfo {author} {\bibfnamefont {J.~F.}\ \bibnamefont {Clauser}}\ and\ \bibinfo {author} {\bibfnamefont {A.}~\bibnamefont {Shimony}},\ }\bibfield  {title} {\bibinfo {title} {Bell's theorem. {{Experimental}} tests and implications},\ }\href {https://doi.org/10.1088/0034-4885/41/12/002} {\bibfield  {journal} {\bibinfo  {journal} {Rep. Prog. Phys.}\ }\textbf {\bibinfo {volume} {41}},\ \bibinfo {pages} {1881} (\bibinfo {year} {1978})}\BibitemShut {NoStop}%
\bibitem [{\citenamefont {{Lewis-Swan}}(2016)}]{lewis-swanUltracold2016}%
  \BibitemOpen
  \bibfield  {author} {\bibinfo {author} {\bibfnamefont {R.~J.}\ \bibnamefont {{Lewis-Swan}}},\ }\href@noop {} {\emph {\bibinfo {title} {Ultracold {{Atoms}} for {{Foundational Tests}} of {{Quantum Mechanics}}}}},\ Springer {{Theses}}\ (\bibinfo  {publisher} {Springer International Publishing},\ \bibinfo {address} {Cham},\ \bibinfo {year} {2016})\BibitemShut {NoStop}%
\bibitem [{\citenamefont {Wiseman}\ \emph {et~al.}(2007)\citenamefont {Wiseman}, \citenamefont {Jones},\ and\ \citenamefont {Doherty}}]{wisemanSteering2007a}%
  \BibitemOpen
  \bibfield  {author} {\bibinfo {author} {\bibfnamefont {H.~M.}\ \bibnamefont {Wiseman}}, \bibinfo {author} {\bibfnamefont {S.~J.}\ \bibnamefont {Jones}},\ and\ \bibinfo {author} {\bibfnamefont {A.~C.}\ \bibnamefont {Doherty}},\ }\bibfield  {title} {\bibinfo {title} {Steering, {{Entanglement}}, {{Nonlocality}}, and the {{Einstein-Podolsky-Rosen Paradox}}},\ }\href {https://doi.org/10.1103/PhysRevLett.98.140402} {\bibfield  {journal} {\bibinfo  {journal} {Phys. Rev. Lett.}\ }\textbf {\bibinfo {volume} {98}},\ \bibinfo {pages} {140402} (\bibinfo {year} {2007})}\BibitemShut {NoStop}%
\bibitem [{\citenamefont {Cavalcanti}\ \emph {et~al.}(2009)\citenamefont {Cavalcanti}, \citenamefont {Jones}, \citenamefont {Wiseman},\ and\ \citenamefont {Reid}}]{cavalcantiExperimental2009a}%
  \BibitemOpen
  \bibfield  {author} {\bibinfo {author} {\bibfnamefont {E.~G.}\ \bibnamefont {Cavalcanti}}, \bibinfo {author} {\bibfnamefont {S.~J.}\ \bibnamefont {Jones}}, \bibinfo {author} {\bibfnamefont {H.~M.}\ \bibnamefont {Wiseman}},\ and\ \bibinfo {author} {\bibfnamefont {M.~D.}\ \bibnamefont {Reid}},\ }\bibfield  {title} {\bibinfo {title} {Experimental criteria for steering and the {{Einstein-Podolsky-Rosen}} paradox},\ }\href {https://doi.org/10.1103/PhysRevA.80.032112} {\bibfield  {journal} {\bibinfo  {journal} {Phys. Rev. A}\ }\textbf {\bibinfo {volume} {80}},\ \bibinfo {pages} {032112} (\bibinfo {year} {2009})}\BibitemShut {NoStop}%
\bibitem [{\citenamefont {Yan}\ \emph {et~al.}(2024)\citenamefont {Yan}, \citenamefont {Kannan}, \citenamefont {Athreya}, \citenamefont {Truscott},\ and\ \citenamefont {Hodgman}}]{yanProposal2024a}%
  \BibitemOpen
  \bibfield  {author} {\bibinfo {author} {\bibfnamefont {X.~T.}\ \bibnamefont {Yan}}, \bibinfo {author} {\bibfnamefont {S.}~\bibnamefont {Kannan}}, \bibinfo {author} {\bibfnamefont {Y.~S.}\ \bibnamefont {Athreya}}, \bibinfo {author} {\bibfnamefont {A.~G.}\ \bibnamefont {Truscott}},\ and\ \bibinfo {author} {\bibfnamefont {S.~S.}\ \bibnamefont {Hodgman}},\ }\bibfield  {title} {\bibinfo {title} {Proposal for a {{Bell}} test with entangled atoms of different mass},\ }\href {https://doi.org/10.48550/arXiv.2411.08356} {\bibfield  {journal} {\bibinfo  {journal} {arXiv:2411.08356}\ } (\bibinfo {year} {2024})}\BibitemShut {NoStop}%
\bibitem [{\citenamefont {Geiger}\ and\ \citenamefont {Trupke}(2018)}]{geigerProposal2018a}%
  \BibitemOpen
  \bibfield  {author} {\bibinfo {author} {\bibfnamefont {R.}~\bibnamefont {Geiger}}\ and\ \bibinfo {author} {\bibfnamefont {M.}~\bibnamefont {Trupke}},\ }\bibfield  {title} {\bibinfo {title} {Proposal for a {{Quantum Test}} of the {{Weak Equivalence Principle}} with {{Entangled Atomic Species}}},\ }\href {https://doi.org/10.1103/PhysRevLett.120.043602} {\bibfield  {journal} {\bibinfo  {journal} {Phys. Rev. Lett.}\ }\textbf {\bibinfo {volume} {120}},\ \bibinfo {pages} {043602} (\bibinfo {year} {2018})}\BibitemShut {NoStop}%
\bibitem [{\citenamefont {Alvarez}(1989)}]{alvarezQuantum1989}%
  \BibitemOpen
  \bibfield  {author} {\bibinfo {author} {\bibfnamefont {E.}~\bibnamefont {Alvarez}},\ }\bibfield  {title} {\bibinfo {title} {Quantum gravity: An introduction to some recent results},\ }\href {https://doi.org/10.1103/RevModPhys.61.561} {\bibfield  {journal} {\bibinfo  {journal} {Rev. Mod. Phys.}\ }\textbf {\bibinfo {volume} {61}},\ \bibinfo {pages} {561} (\bibinfo {year} {1989})}\BibitemShut {NoStop}%
\bibitem [{\citenamefont {Zych}\ \emph {et~al.}(2011)\citenamefont {Zych}, \citenamefont {Costa}, \citenamefont {Pikovski},\ and\ \citenamefont {Brukner}}]{zychQuantum2011}%
  \BibitemOpen
  \bibfield  {author} {\bibinfo {author} {\bibfnamefont {M.}~\bibnamefont {Zych}}, \bibinfo {author} {\bibfnamefont {F.}~\bibnamefont {Costa}}, \bibinfo {author} {\bibfnamefont {I.}~\bibnamefont {Pikovski}},\ and\ \bibinfo {author} {\bibfnamefont {{\v C}.}~\bibnamefont {Brukner}},\ }\bibfield  {title} {\bibinfo {title} {Quantum interferometric visibility as a witness of general relativistic proper time},\ }\href {https://doi.org/10.1038/ncomms1498} {\bibfield  {journal} {\bibinfo  {journal} {Nat. Commun.}\ }\textbf {\bibinfo {volume} {2}},\ \bibinfo {pages} {505} (\bibinfo {year} {2011})}\BibitemShut {NoStop}%
\bibitem [{\citenamefont {Vowe}\ \emph {et~al.}(2022)\citenamefont {Vowe}, \citenamefont {Donadi}, \citenamefont {Schkolnik}, \citenamefont {Peters}, \citenamefont {Leykauf},\ and\ \citenamefont {Krutzik}}]{voweLightpulseAtomInterferometric2022}%
  \BibitemOpen
  \bibfield  {author} {\bibinfo {author} {\bibfnamefont {S.}~\bibnamefont {Vowe}}, \bibinfo {author} {\bibfnamefont {S.}~\bibnamefont {Donadi}}, \bibinfo {author} {\bibfnamefont {V.}~\bibnamefont {Schkolnik}}, \bibinfo {author} {\bibfnamefont {A.}~\bibnamefont {Peters}}, \bibinfo {author} {\bibfnamefont {B.}~\bibnamefont {Leykauf}},\ and\ \bibinfo {author} {\bibfnamefont {M.}~\bibnamefont {Krutzik}},\ }\bibfield  {title} {\bibinfo {title} {Light-pulse atom interferometric test of continuous spontaneous localization},\ }\href {https://doi.org/10.1103/PhysRevA.106.043317} {\bibfield  {journal} {\bibinfo  {journal} {Phys. Rev. A}\ }\textbf {\bibinfo {volume} {106}},\ \bibinfo {pages} {043317} (\bibinfo {year} {2022})}\BibitemShut {NoStop}%
\bibitem [{\citenamefont {Braunstein}\ and\ \citenamefont {van Loock}(2005)}]{braunsteinQuantumInformationContinuous2005}%
  \BibitemOpen
  \bibfield  {author} {\bibinfo {author} {\bibfnamefont {S.~L.}\ \bibnamefont {Braunstein}}\ and\ \bibinfo {author} {\bibfnamefont {P.}~\bibnamefont {van Loock}},\ }\bibfield  {title} {\bibinfo {title} {Quantum information with continuous variables},\ }\href {https://doi.org/10.1103/RevModPhys.77.513} {\bibfield  {journal} {\bibinfo  {journal} {Rev. Mod. Phys.}\ }\textbf {\bibinfo {volume} {77}},\ \bibinfo {pages} {513} (\bibinfo {year} {2005})}\BibitemShut {NoStop}%
\bibitem [{\citenamefont {O'Sullivan-Hale}\ \emph {et~al.}(2005)\citenamefont {O'Sullivan-Hale}, \citenamefont {Ali~Khan}, \citenamefont {Boyd},\ and\ \citenamefont {Howell}}]{OSullivanHalePixelEntanglement2005}%
  \BibitemOpen
  \bibfield  {author} {\bibinfo {author} {\bibfnamefont {M.~N.}\ \bibnamefont {O'Sullivan-Hale}}, \bibinfo {author} {\bibfnamefont {I.}~\bibnamefont {Ali~Khan}}, \bibinfo {author} {\bibfnamefont {R.~W.}\ \bibnamefont {Boyd}},\ and\ \bibinfo {author} {\bibfnamefont {J.~C.}\ \bibnamefont {Howell}},\ }\bibfield  {title} {\bibinfo {title} {Pixel entanglement: Experimental realization of optically entangled $d=3$ and $d=6$ qudits},\ }\href {https://doi.org/10.1103/PhysRevLett.94.220501} {\bibfield  {journal} {\bibinfo  {journal} {Phys. Rev. Lett.}\ }\textbf {\bibinfo {volume} {94}},\ \bibinfo {pages} {220501} (\bibinfo {year} {2005})}\BibitemShut {NoStop}%
\bibitem [{\citenamefont {Walborn}\ \emph {et~al.}(2006)\citenamefont {Walborn}, \citenamefont {Lemelle}, \citenamefont {Almeida},\ and\ \citenamefont {Ribeiro}}]{WalbornQKDSpatialQubits2006}%
  \BibitemOpen
  \bibfield  {author} {\bibinfo {author} {\bibfnamefont {S.~P.}\ \bibnamefont {Walborn}}, \bibinfo {author} {\bibfnamefont {D.~S.}\ \bibnamefont {Lemelle}}, \bibinfo {author} {\bibfnamefont {M.~P.}\ \bibnamefont {Almeida}},\ and\ \bibinfo {author} {\bibfnamefont {P.~H.~S.}\ \bibnamefont {Ribeiro}},\ }\bibfield  {title} {\bibinfo {title} {Quantum key distribution with higher-order alphabets using spatially encoded qudits},\ }\href {https://doi.org/10.1103/PhysRevLett.96.090501} {\bibfield  {journal} {\bibinfo  {journal} {Phys. Rev. Lett.}\ }\textbf {\bibinfo {volume} {96}},\ \bibinfo {pages} {090501} (\bibinfo {year} {2006})}\BibitemShut {NoStop}%
\bibitem [{\citenamefont {Anders}\ \emph {et~al.}(2021)\citenamefont {Anders}, \citenamefont {Idel}, \citenamefont {Feldmann}, \citenamefont {Bondarenko}, \citenamefont {Loriani}, \citenamefont {Lange}, \citenamefont {Peise}, \citenamefont {Gersemann}, \citenamefont {{Meyer-Hoppe}}, \citenamefont {Abend}, \citenamefont {Gaaloul}, \citenamefont {Schubert}, \citenamefont {Schlippert}, \citenamefont {Santos}, \citenamefont {Rasel},\ and\ \citenamefont {Klempt}}]{anders2021MomentumEntanglementAtom}%
  \BibitemOpen
  \bibfield  {author} {\bibinfo {author} {\bibfnamefont {F.}~\bibnamefont {Anders}}, \bibinfo {author} {\bibfnamefont {A.}~\bibnamefont {Idel}}, \bibinfo {author} {\bibfnamefont {P.}~\bibnamefont {Feldmann}}, \bibinfo {author} {\bibfnamefont {D.}~\bibnamefont {Bondarenko}}, \bibinfo {author} {\bibfnamefont {S.}~\bibnamefont {Loriani}}, \bibinfo {author} {\bibfnamefont {K.}~\bibnamefont {Lange}}, \bibinfo {author} {\bibfnamefont {J.}~\bibnamefont {Peise}}, \bibinfo {author} {\bibfnamefont {M.}~\bibnamefont {Gersemann}}, \bibinfo {author} {\bibfnamefont {B.}~\bibnamefont {{Meyer-Hoppe}}}, \bibinfo {author} {\bibfnamefont {S.}~\bibnamefont {Abend}}, \bibinfo {author} {\bibfnamefont {N.}~\bibnamefont {Gaaloul}}, \bibinfo {author} {\bibfnamefont {C.}~\bibnamefont {Schubert}}, \bibinfo {author} {\bibfnamefont {D.}~\bibnamefont {Schlippert}}, \bibinfo {author} {\bibfnamefont {L.}~\bibnamefont {Santos}}, \bibinfo {author} {\bibfnamefont {E.}~\bibnamefont {Rasel}},\ and\ \bibinfo {author} {\bibfnamefont
  {C.}~\bibnamefont {Klempt}},\ }\bibfield  {title} {\bibinfo {title} {Momentum {{Entanglement}} for {{Atom Interferometry}}},\ }\href {https://doi.org/10.1103/PhysRevLett.127.140402} {\bibfield  {journal} {\bibinfo  {journal} {Phys. Rev. Lett.}\ }\textbf {\bibinfo {volume} {127}},\ \bibinfo {pages} {140402} (\bibinfo {year} {2021})}\BibitemShut {NoStop}%
\bibitem [{\citenamefont {Fadel}\ \emph {et~al.}(2024)\citenamefont {Fadel}, \citenamefont {Roux},\ and\ \citenamefont {Gessner}}]{fadelQuantumMetrologyContinuousvariable2024}%
  \BibitemOpen
  \bibfield  {author} {\bibinfo {author} {\bibfnamefont {M.}~\bibnamefont {Fadel}}, \bibinfo {author} {\bibfnamefont {N.}~\bibnamefont {Roux}},\ and\ \bibinfo {author} {\bibfnamefont {M.}~\bibnamefont {Gessner}},\ }\bibfield  {title} {\bibinfo {title} {Quantum metrology with a continuous-variable system},\ }\href {https://doi.org/10.48550/arXiv.2411.04122} {\bibfield  {journal} {\bibinfo  {journal} {arXiv:2411.04122}\ } (\bibinfo {year} {2024})}\BibitemShut {NoStop}%
\bibitem [{\citenamefont {Kannan}\ \emph {et~al.}(2024)\citenamefont {Kannan}, \citenamefont {Athreya}, \citenamefont {Abbas}, \citenamefont {Yan}, \citenamefont {Hodgman},\ and\ \citenamefont {Truscott}}]{kannanMeasurement2024}%
  \BibitemOpen
  \bibfield  {author} {\bibinfo {author} {\bibfnamefont {S.}~\bibnamefont {Kannan}}, \bibinfo {author} {\bibfnamefont {Y.~S.}\ \bibnamefont {Athreya}}, \bibinfo {author} {\bibfnamefont {A.~H.}\ \bibnamefont {Abbas}}, \bibinfo {author} {\bibfnamefont {X.~T.}\ \bibnamefont {Yan}}, \bibinfo {author} {\bibfnamefont {S.~S.}\ \bibnamefont {Hodgman}},\ and\ \bibinfo {author} {\bibfnamefont {A.~G.}\ \bibnamefont {Truscott}},\ }\bibfield  {title} {\bibinfo {title} {Measurement of the s -wave scattering length between metastable helium isotopes},\ }\href {https://doi.org/10.1103/PhysRevA.110.063324} {\bibfield  {journal} {\bibinfo  {journal} {Phys. Rev. A}\ }\textbf {\bibinfo {volume} {110}},\ \bibinfo {pages} {063324} (\bibinfo {year} {2024})}\BibitemShut {NoStop}%
\bibitem [{\citenamefont {Henson}\ \emph {et~al.}(2018)\citenamefont {Henson}, \citenamefont {Yue}, \citenamefont {Hodgman}, \citenamefont {Shin}, \citenamefont {Smirnov}, \citenamefont {Ostrovskaya}, \citenamefont {Guan},\ and\ \citenamefont {Truscott}}]{hensonBogoliubovCherenkov2018}%
  \BibitemOpen
  \bibfield  {author} {\bibinfo {author} {\bibfnamefont {B.~M.}\ \bibnamefont {Henson}}, \bibinfo {author} {\bibfnamefont {X.}~\bibnamefont {Yue}}, \bibinfo {author} {\bibfnamefont {S.~S.}\ \bibnamefont {Hodgman}}, \bibinfo {author} {\bibfnamefont {D.~K.}\ \bibnamefont {Shin}}, \bibinfo {author} {\bibfnamefont {L.~A.}\ \bibnamefont {Smirnov}}, \bibinfo {author} {\bibfnamefont {E.~A.}\ \bibnamefont {Ostrovskaya}}, \bibinfo {author} {\bibfnamefont {X.~W.}\ \bibnamefont {Guan}},\ and\ \bibinfo {author} {\bibfnamefont {A.~G.}\ \bibnamefont {Truscott}},\ }\bibfield  {title} {\bibinfo {title} {Bogoliubov-{{Cherenkov}} radiation in an atom laser},\ }\href {https://doi.org/10.1103/PhysRevA.97.063601} {\bibfield  {journal} {\bibinfo  {journal} {Phys. Rev. A}\ }\textbf {\bibinfo {volume} {97}},\ \bibinfo {pages} {063601} (\bibinfo {year} {2018})}\BibitemShut {NoStop}%
\bibitem [{\citenamefont {Cetinkaya-Rundel}\ \emph {et~al.}(2019)\citenamefont {Cetinkaya-Rundel}, \citenamefont {Diez},\ and\ \citenamefont {Barr}}]{Cetinkaya-RundelOpenIntroStat2019}%
  \BibitemOpen
  \bibfield  {author} {\bibinfo {author} {\bibfnamefont {M.}~\bibnamefont {Cetinkaya-Rundel}}, \bibinfo {author} {\bibfnamefont {D.}~\bibnamefont {Diez}},\ and\ \bibinfo {author} {\bibfnamefont {C.}~\bibnamefont {Barr}},\ }\href@noop {} {\emph {\bibinfo {title} {OpenIntro Statistics}}},\ \bibinfo {edition} {{{Fourth}} edition}\ ed.\ (\bibinfo  {publisher} {OpenIntro, Inc.},\ \bibinfo {year} {2019})\BibitemShut {NoStop}%
\bibitem [{\citenamefont {Cavalcanti}\ \emph {et~al.}(2015)\citenamefont {Cavalcanti}, \citenamefont {Foster}, \citenamefont {Fuwa},\ and\ \citenamefont {Wiseman}}]{cavalcantiAnalog2015}%
  \BibitemOpen
  \bibfield  {author} {\bibinfo {author} {\bibfnamefont {E.~G.}\ \bibnamefont {Cavalcanti}}, \bibinfo {author} {\bibfnamefont {C.~J.}\ \bibnamefont {Foster}}, \bibinfo {author} {\bibfnamefont {M.}~\bibnamefont {Fuwa}},\ and\ \bibinfo {author} {\bibfnamefont {H.~M.}\ \bibnamefont {Wiseman}},\ }\bibfield  {title} {\bibinfo {title} {Analog of the {{Clauser}}--{{Horne}}--{{Shimony}}--{{Holt}} inequality for steering},\ }\href {https://doi.org/10.1364/JOSAB.32.000A74} {\bibfield  {journal} {\bibinfo  {journal} {JOSA B}\ }\textbf {\bibinfo {volume} {32}},\ \bibinfo {pages} {A74} (\bibinfo {year} {2015})}\BibitemShut {NoStop}%
\end{thebibliography}%

%%%%%%%%%%%%%% Appending supplementary %%%%%%%%%%%%%
\clearpage
\onecolumngrid
\begin{center}
\textbf{{\mdseries\large Supplementary material for}\\ \large Bell correlations between momentum-entangled pairs of \textsuperscript{4}He\textsuperscript{*} atoms}
\vspace{1cm}
\end{center}

\setcounter{equation}{0}
\setcounter{figure}{0}
\setcounter{table}{0}
\setcounter{section}{0}
\renewcommand{\figurename}{\textbf{Fig.}}
\renewcommand{\thefigure}{\textbf{S}\textbf{\arabic{figure}}}
\renewcommand{\theequation}{\text{S.}\arabic{equation}}

%%%%%%%%%%%%%%%%%%% Appending supplementary %%%%%%%%%%%%

%-----------------------------------------------------------------
\section{\label{Bragg} Bragg pulse characterization}
\begin{figure}[h]
    \includegraphics[scale=0.75]{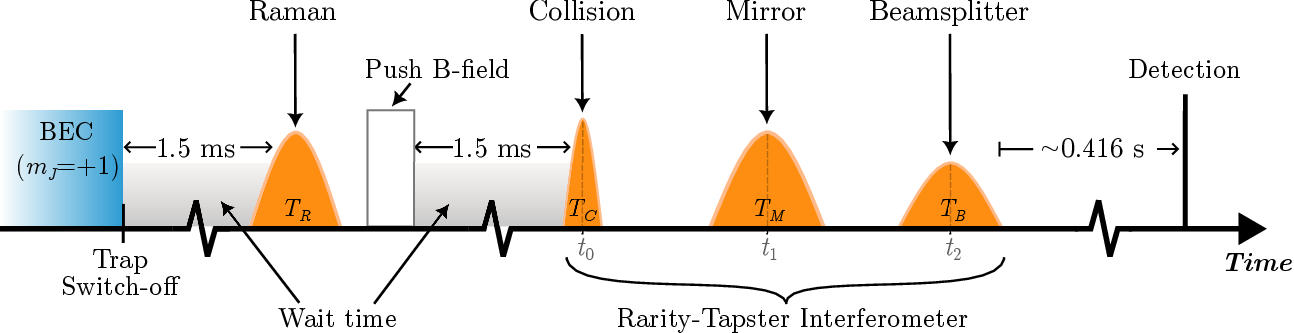}
    \caption{\textbf{Schematic of experimental sequence}. Following the formation of the $^4\text{He}^*$ BEC in the $m_J=+1$ sublevel, the magnetic trap holding the atoms is rapidly switched off. We then wait 1.5 ms to allow the magnetic field to stabilize to a uniform field $\textbf{B}_0 \approx [0.5(\hat{\mathbf{x}} + \hat{\mathbf{z}}) - 0.8\hat{\mathbf{y}}] $ G set and actively stabilised by 6 external magnetic field compensation coils \cite{dallBose2007a}. A two-photon Raman transition pulse is then applied for a duration $T_R=5.6$ $\mu$s to transfer the atoms to the magnetically insensitive $m_J=0$ state. We eliminate any non-transferred $m_J=1$ atoms by applying a `push B-field' (a magnetic field gradient) to expel these atoms outside the detection range. This ensures that the atoms undergoing the interferometric experiment are unaffected by stray magnetic fields that may be present in the vacuum chamber. At $t_0$, we begin the Rarity-Tapster interferometric sequence by applying a series of Bragg pulses - collision($t_0$), mirror($t_1$) and beamsplitter($t_2$) with pulse durations $T_C$, $T_M$ and $T_B$, respectively. At the end of the interferometric scheme, the scattered atoms fall for  $t_\text{f}\approx 0.416$ s onto the MCP-DLD setup for detection.}
    \label{fig:supplfig1}
\end{figure}

\begin{figure}
    \centering
    \includegraphics[width=\linewidth]{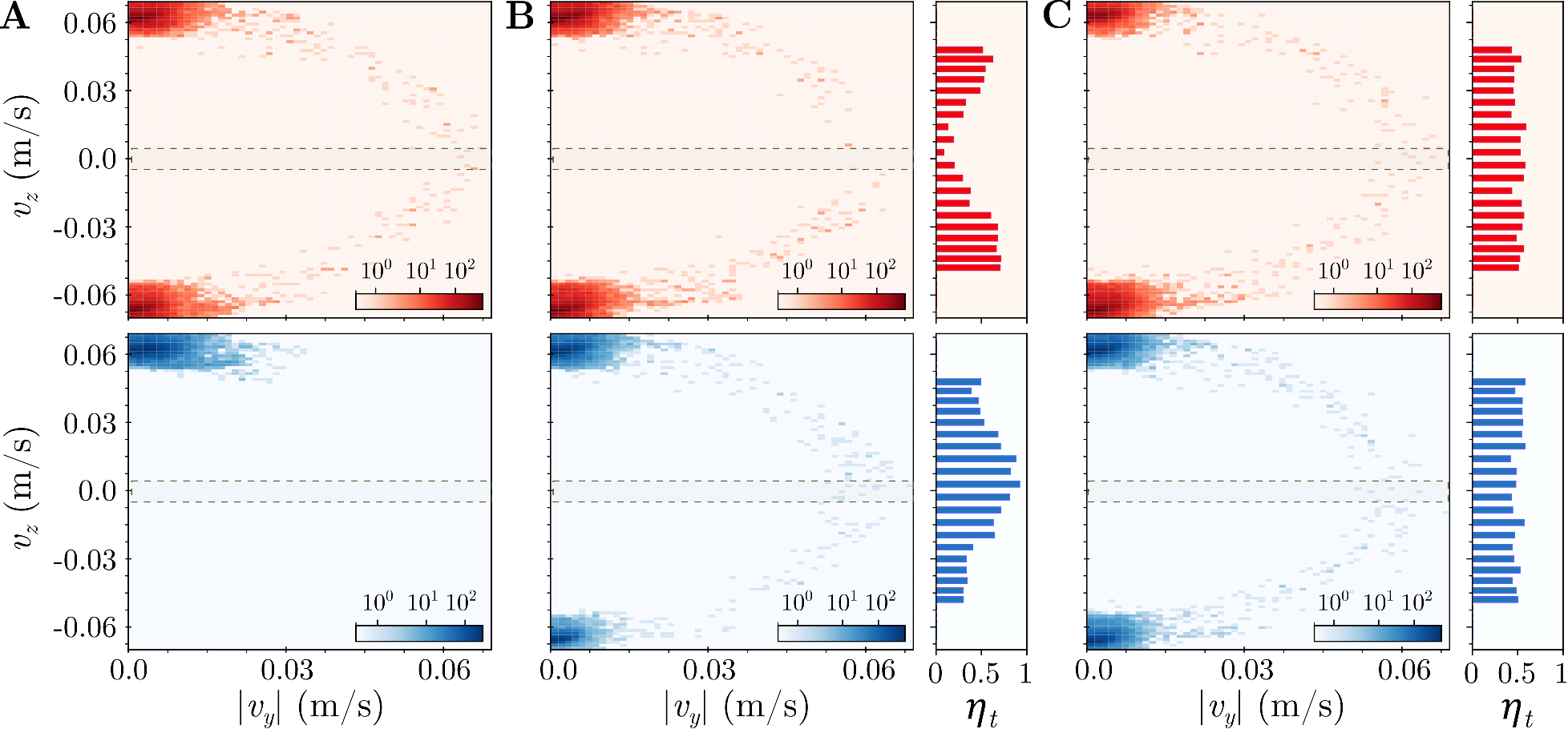}
    \caption{
    \textbf{Bragg pulse transfer optimisation}. 2D histogram plots, integrated over 50 shots, illustrate a slice of the detected counts for the top (red) and bottom (blue) halos in their respective center-of-velocity reference frames. The grey-shaded boxes indicate regions of interest around the halo equators, corresponding to the detection windows used in the experiment. \textbf{(A)} A top halo is generated before optimal Bragg pulses are applied. \textbf{(B)}
    The mirror pulse reflects atoms across the top halo equator to the bottom halo, achieving a peak transfer efficiency of $90(5)\%$. \textbf{(C)} While the beamsplitter pulse yields an average transfer efficiency of $50(5)\%$. The bar graphs to the right of the 2D histogram plots display the corresponding transfer efficiencies ($\eta_t$) for the respective pulse types in \textbf{(B)} and \textbf{(C)}. Similar transfer efficiencies are also observed when these pulses are applied to an initially generated bottom halo. 
    }
    \label{fig:supplfig2}
\end{figure}

The experiment described in the main text consists of a series of light pulses to transfer various groups of atoms between different momentum and internal states. This pulse sequence is shown in Fig.~\ref{fig:supplfig1}. In the Rarity-Tapster interferometer scheme, we utilize Bragg transition pulses, starting with a collision pulse ($t_0=0$) with duration $T_{C} = 2.4$ $\mu$s. Following this we apply a $\pi$-Bragg pulse or mirror pulse at $t_1=350$ $\mu$s for $T_{M}=7$ $\mu$s to reflect selected scattering modes in the halos onto each other. Due to our laser geometry setup and the use of a single set of Bragg beams in our experiment, we select scattering pairs about the equator of the halos. This selection of the scattering modes allows us to use the same set of Bragg beams to couple counter-propagating (moving $\pm\hbar \text{k}_0 \hat{\textbf{z}}$ in relation to the source $m_J=0$ cloud of atoms) scattering modes of ($\textbf{p},\textbf{p}^\prime$) and ($\textbf{q},\textbf{q}^\prime$), as shown in Fig.~1C of main text.\\
We optimize the Gaussian-modulated pulse amplitude, duration, detuning and timing of the mirror pulse to achieve maximum coupling efficiency resonant to our selected modes about the halo equator (i.e., $\textit{v}_z = 0$ m/s in the centre-of-momentum reference frame of the halo). All optimisation is performed using a single halo (Fig.~\ref{fig:supplfig2}A), where only either ($\textbf{p},\textbf{p}^\prime$) or ($\textbf{q},\textbf{q}^\prime$) are initially diffracted, and then a mirror pulse is applied to achieve maximum transfer to its counter-propagating pair, i.e., ($\textbf{p},\textbf{p}^\prime$) $\rightarrow$ ($\textbf{q},\textbf{q}^\prime$) or ($\textbf{q},\textbf{q}^\prime$) $\rightarrow$ ($\textbf{p},\textbf{p}^\prime$). The results of this optimization procedure are shown in Fig.~\ref{fig:supplfig2}B. Similarly, for the beamsplitter pulse, we apply a $\pi/2$-Bragg pulse at $t_2=700$ $\mu$s for $T_{B}=6.2$ $\mu$s and optimize the pulse as described above. The results of the beamsplitter optimization are shown in Fig.~\ref{fig:supplfig2}C.\\ 

%-----------------------------------------------------------------------
\section{\label{Analysis} Data Analysis and post-selection}

Exploiting the $\sim$19.8 eV internal energy of the metastable helium atoms, we can measure single atom detection events at the microchannel plate (MCP) and delay line detector (DLD) with an estimated quantum efficiency of 20(2)$\%$ \cite{kannanMeasurement2024} and spatial-temporal resolutions of 120 $\mu$m and 3 $\mu$s \cite{hensonBogoliubovCherenkov2018}, corresponding to a momentum resolution of about $\sim(4.5\times10^{-3})k_0$ along $k_{x,y}$ and $\sim(4.6\times10^{-4})k_0$ along $k_z$. 

From the reconstructed velocities of the atoms, we accurately determine the momenta of the scattered atoms by performing coordinate transforms for each of the halos into the centre-of-momentum (COM) reference frame for respective halos. Due to the number of scattered atoms in the halos being relatively few ($<100$), we determine the COM of the halo using its colliding parent condensates which have a higher number of atoms ($\sim3\times10^3$). This estimation of the position of the COM is a good approximation of the true COM of the halo. Further, we perform this coordinate transform to the halo COM reference frame for each experimental run, thus minimizing any broadening or skewing of momentum distributions due to shot-to-shot fluctuations when integrating over multiple datasets. We also perform dataset filtering based on the number of scattered atoms in the halos to ensure that we analyse datasets having similar mode occupancy ($\bar{n}\approx0.035$) and hence consistent two-particle correlation amplitudes $g^{(2)}(0)$.

We obtain the measured correlation results by limiting our detection windows around each of the halo equators. This is to ensure that selected momentum modes within our detection window experience nearly uniform and identical coupling from the Bragg transfer pulses [Suppl. section \ref{Bragg}]. Specifically, we only include scattered pairs having radial velocities of $0.8\leq v/v_r \leq 1.1$  ($v_r$ is the velocity radius of the scattering halo equal to $\sim65$ mm/s), and pairs within $\pm4^\circ$ from the halo's equatorial plane, corresponding to a vertical velocity range of $\pm4.5$ mm/s about the equator. We arrive at this detection window range as a balance between two competing constraints – minimizing unintentional averaging of the imprinted phases of the scattered pairs across the detection range resulting in decreased interference contrast and reduced joint-detection probabilities at the output of the interferometer, and maximizing the signal-to-noise ratio (SNR) through integrating over a wider scattered momentum range. Our chosen detection window ranges \{$D_{L1},D_{L2},D_{R1},D_{R2}$\} are shown in Fig.~1C, within which the phase varies by approximately 1.5 radians due to particle-path trajectories, and we achieve an average SNR of 30:1. From this truncated distribution we measure the pair-correlations between the scattered atoms, where we define the size of the integration volume (equivalently, size of a scattering mode) as a cubic volume with sides equal to $\sim0.02\times\text{k}_r$ ($\text{k}_r = m\cdot v_r/\hbar$). The side length of integration volume is approximately $0.5\sigma_{BB}$, where $\sigma_{BB}$ represents the two-atom back-to-back correlation length \cite{hodgmanSolving2017}. This length follows from the relation $\sigma_{BB}\approx 1.1\sigma_k $, with $\sigma_k$ being the rms momentum width of the source condensate, assuming it can be approximated by an isotropic Gaussian distribution \cite{ogrenAtomatom2009}. The small integration volume size also ensures several bins are within the detection window range.

%----------------------------------------------------------------------
\section{\label{Error}Error Analysis}
By selecting experimental runs where we only observe a single scattered pair in our detection window ranges, in each individual experimental run we only ever obtain a measured value of the Bell correlation function $E$ (Eq.\,(5) of the main text) to be either  $1$ or $-1$. Given the number of unique values in the sample space of $E$ is only two, we use a binomial proportional estimator \cite{Cetinkaya-RundelOpenIntroStat2019}, derived from the binomial distribution, in our statistical analysis. Error bars for a binomial proportion are calculated using the standard error (SE) of the proportion, which is $\text{SE} = \sqrt{\frac{p(1 - p)}{n}}$, where $p$ is the proportion of successes ($1$ or $-1$) and $n$ is the sample size. 
Identical binary set outcome values are also observed in the measurement of the joint probability distribution function $P_{\textbf{k},\textbf{k}^\prime}$ from post-selected runs and thus supporting the use of a binomial proportion estimator. 

%-----------------------------------------------------------------------
\section{\label{Bell state}Two-particle momentum-entangled Bell state}
The preparation of the Bell state in our experiment is realised through atom-atom interactions where pairwise scattering between individual atoms in colliding condensates gives rise to entangled pairs of atoms being spontaneously emitted into distinguishable momentum modes. The derivation of this mapping follows from \cite{thomasMatterwave2022,lewis-swanProposal2015}. The collision of atomic ensembles can be understood as the spontaneous four-wave mixing (SFWM) of matter-waves and their collisional non-linear interaction, which creates entangled pairs of atoms with opposite correlated momenta. We begin by considering a single halo where we describe the collision mechanism between colliding pairs of condensates using a sum of the two-mode squeezing Hamiltonian $\hat{H}=\sum_{\textbf{k}}{\hbar\zeta(\hat{a}_{\textbf{k}}^\dagger\hat{a}_{\textbf{k}^\prime}^\dagger+\hat{a}_{\textbf{k}^\prime}\hat{a}_{\textbf{k}})}$, where $\textbf{k}+\textbf{k}^\prime = 2\textbf{k}_0$, $\hat{a}_{\textbf{k}(\textbf{k}^\prime)}^{\dagger}$ is the bosonic creation operator acting on momentum mode $\textbf{k}(\textbf{k}^\prime)$, and $\zeta$ is an effective nonlinearity factor dependent on the colliding species. Assuming an initial vacuum state for all momenta (excluding the momenta of the colliding condensate pair), the squeezing Hamiltonian generates a product of two-mode squeezed vacuum states (TMSV) describing the scattering halo as
\begin{equation}
\ket{\psi}_{\text{TMSV}\text{, halo}} = \prod_{\textbf{k}+\textbf{k}^\prime=2\textbf{k}_0}\otimes \left( \sqrt{1-\mu^2}\sum_{n=0}^{\infty}{\mu^n\ket{n}_{\textbf{k}}\ket{n}_{\textbf{k}^\prime}}\right),
\label{eq:S1}
\end{equation}
where $\mu = \tanh(\zeta t_c)$ for a collision duration $t_c$. This state is a product of number-correlated superposition states $\ket{n}_{\textbf{k}}\ket{n}_{\textbf{k}^\prime}$ across the halo satisfying $\textbf{k}+\textbf{k}^\prime = 2\textbf{k}_0$. Similarly, as described by Eq.\,\eqref{eq:S1}, the lower halo can be equivalently constructed with the condition $\textbf{k}+\textbf{k}^\prime=-2\textbf{k}_0$.   From this, we formulate our initial double $s$-wave collision halo configuration as a product state between the independent upper and lower halo states, assuming identical collision parameters and negligible depletion of the source condensate: $\ket{\Psi}_{\text{double halo}} = \ket{\psi}_{\text{upper halo}}\otimes \ket{\psi}_{\text{lower halo}}$. We further simplify the double halo state $\ket{\Psi}_{\text{double halo}}$ by tracing away momentum modes not equal to our selected momentum modes ($\textbf{p},\textbf{p}^\prime,\textbf{q},\textbf{q}^\prime$) at the halo equators. This results in the following four-mode state for our selected momentum modes in the double halo:
\begin{equation}
    \ket{\Psi}=(1-\mu^2)\sum_{n,m=0}^{\infty}{\mu^{n+m}\ket{n}_{\textbf{p}}\ket{n}_{\textbf{p}^\prime}\ket{m}_{\textbf{q}}\ket{m}_{\textbf{q}^\prime}}.
    \label{S2}
\end{equation}
\\
In the low gain, perturbative regime, $\mu\ll1$ and thus $\bar{n}=\frac{\mu^2}{1-\mu^2}\ll1$. This allows us to truncate Eq.\,\eqref{S2} to the first order in $\mu$, which corresponds to ignoring contributions from scattering of multiple pairs - i.e., Fock states with $n,m > 1$. We truncate the above scattering state to consider only occupancies $\leq2$ particles across the two halos (four modes), resulting in the following truncated wavefunction:
\begin{equation}
    \ket{\Psi}\approx(1-\mu^2)\left(\ket{0}_{\textbf{p}}\ket{0}_{\textbf{p}^\prime}\ket{0}_{\textbf{q}}\ket{0}_{\textbf{q}^\prime}+\mu\ket{0}_{\textbf{p}}\ket{0}_{\textbf{p}^\prime}\ket{1}_{\textbf{q}}\ket{1}_{\textbf{q}^\prime}+\mu\ket{1}_{\textbf{p}}\ket{1}_{\textbf{p}^\prime}\ket{0}_{\textbf{q}}\ket{0}_{\textbf{q}^\prime} \right).
    \label{S3}
\end{equation}
As the vacuum state $\ket{0}_{\textbf{p}}\ket{0}_{\textbf{p}^\prime}\ket{0}_{\textbf{q}}\ket{0}_{\textbf{q}^\prime}$ does not contribute to the experimental correlations we measure, we neglect the first term in Eq.\,\eqref{S3} and further truncate the above expression to
\begin{equation}
        \ket{\Psi}\approx\frac{1}{\sqrt{2}}\left(\ket{0}_{\textbf{p}}\ket{0}_{\textbf{p}^\prime}\ket{1}_{\textbf{q}}\ket{1}_{\textbf{q}^\prime}+\ket{1}_{\textbf{p}}\ket{1}_{\textbf{p}^\prime}\ket{0}_{\textbf{q}}\ket{0}_{\textbf{q}^\prime} \right),
    \label{S3a}
\end{equation}
where we have enforced normalisation. This non-separable two-particle state is entangled between atom pairs of selected momentum modes ($\textbf{p},\textbf{p}^\prime$) and ($\textbf{q},\textbf{q}^\prime$). A simple mapping reveals its form is identical to a prototypical Bell state \cite{thomasMatterwave2022,lewis-swanProposal2015}. Experimentally, this truncation is achieved via post-selection of experimental runs that feature only a single pair of atoms within the equatorial modes of interest.

%-----------------------------------------------------------------------
\section{Joint probability distribution function}
\label{Corr func}

Here, we develop an analytical model for the Rarity-Tapster (RT) interferometer setup, assuming ideal Bragg pulses (instantaneous linear transformations) acting on discrete momentum modes with infinite spatial extent. Following similar derivations in \cite{dussarratTwoParticle2017,thomasMatterwave2022}, we can express the coupling Hamiltonian of a Bragg pulse in the basis of \{$\hat{a}_{\textbf{k}},\hat{a}_{\textbf{k}-2\textbf{k}_0}$\} as
\begin{equation}
    \hat{H}=\frac{\hbar\Omega}{2}
    \begin{pmatrix}
        0 & e^{i\phi}\\
        e^{-i\phi} & 0 \\
    \end{pmatrix},
\end{equation}
where $\Omega/2\pi$ is the two-photon Rabi frequency, and $\phi$ is the phase of the Bragg lattice ($=\theta_u - \theta_l$). Using the unitary evolution operator $\hat{U}(t,\phi) = e^{-i\hat{H}t/\hbar}$, we model the Rarity-Tapster setup  as the application of a $\pi$-pulse (mirror) and $\pi/2$-pulse (beamsplitter), i.e.,
\begin{equation}
    \hat{U}(\pi/2\Omega,\phi_{\pi/2})\hat{U}(\pi/\Omega,\phi_\pi)=\hat{U}_{\mathrm{RT}}(\phi_\pi,\phi_\pi/2)=\frac{-1}{\sqrt{2}}
    \begin{pmatrix}
        e^{-i(\phi_{\pi/2}-\phi_\pi)} & ie^{-i\phi_\pi} \\
        ie^{i\phi_\pi} & e^{i(\phi_{\pi/2}-\phi_\pi)}
    \end{pmatrix},
\end{equation}
where $\phi_\pi$ and $\phi_{\pi/2}$ are the phases of the mirror and beamsplitter pulses, respectively. We note here that, due to the unique geometry of our experiment, we realize simultaneous realisations of multiple interferometers coupling resonant momentum modes across the two halos. We express the coupled momenta $\textbf{k}\in\{\textbf{p},\textbf{p}^\prime\}$, $\textbf{k}-2\textbf{k}_0\in\{\textbf{q},\textbf{q}^\prime\}$ and denote the spatially separated interferometric arms as $L$ and $R$, coupling opposing momentum modes in the entangled pairs within each of the halos, i.e., ($\textbf{p},\textbf{q}$) propagates through the $L$ interferometer arm, while ($\textbf{p}^\prime,\textbf{q}^\prime$) propagates through the $R$ interferometer arm (as shown in Fig.~1C).\\
By taking  Eq.\,(1) (the prototypical Bell state) as the initial input state $\ket{\Psi}_\text{in}$  to the RT interferometer, we arrive at the following output state:
\begin{equation}
    \begin{split}
    \ket{\Psi}_{\text{out}}=\hat{U}_{\mathrm{RT}}\ket{\Psi}_\text{in}=\frac{1}{2\sqrt{2}}
        [(1-e^{-i(\phi_L + \phi_R)})\ket{0}_{\textbf{p}}\ket{0}_{\textbf{p}^\prime}\ket{1}_{\textbf{q}}\ket{1}_{\textbf{q}^\prime}\\
        -i(e^{i\phi_R}+e^{-i\phi_L})\ket{0}_{\textbf{p}}\ket{1}_{\textbf{p}^\prime}\ket{1}_{\textbf{q}}\ket{0}_{\textbf{q}^\prime}\\
        -i(e^{i\phi_L}+e^{-i\phi_R})\ket{1}_{\textbf{p}}\ket{0}_{\textbf{p}^\prime}\ket{0}_{\textbf{q}}\ket{1}_{\textbf{q}^\prime}\\
        (1-e^{i(\phi_L + \phi_R)})\ket{1}_{\textbf{p}}\ket{1}_{\textbf{p}^\prime}\ket{0}_{\textbf{q}}\ket{0}_{\textbf{q}^\prime}]
    \end{split}
\end{equation}
where we set the mirror pulse phase ($\phi_\pi$) in both interferometer arms to be $\pi/2$ and represent the phases of the beam splitter pulse ($\phi_{\pi/2}$) for each arm as $\phi_L$ and $\phi_R$.

When measuring correlations between distinct pairs of these modes, we observe a dependence on the applied beamsplitter phase shifts - evidence of interference between scattered pairs. Specifically, we measure the second-order correlation function or joint probability distribution function: $P_{\textbf{k},\textbf{k}^\prime} = \langle\hat{a}^\dagger_{\textbf{k}}\hat{a}^\dagger_{\textbf{k}^\prime}\hat{a}_{\textbf{k}^\prime}\hat{a}_{\textbf{k}}\rangle = \langle \hat{n}_\textbf{k}\hat{n}_{\textbf{k}^\prime} \rangle$ (where $\textbf{k} \in \{\textbf{p},\textbf{q}\}$ and $\textbf{k}^\prime \in \{\textbf{p}^\prime,\textbf{q}^\prime\}$). Evaluating the correlations for $\ket{\Psi}_{\text{out}}$ we obtain,
\begin{equation}
P_{\textbf{p},\textbf{p}^\prime}=P_{\textbf{q},\textbf{q}^\prime} = \frac{1}{2}\sin^2(\frac{\phi_L+\phi_R}{2}),\quad\text{and}\quad
P_{\textbf{p},\textbf{q}^\prime}=P_{\textbf{q},\textbf{p}^\prime} = \frac{1}{2}\cos^2(\frac{\phi_L+\phi_R}{2}).
\label{suppl_eq:joint1}
\end{equation}

Further, if we consider the input state to the interferometer to be the truncated double halo squeezed state (Eq.\,\eqref{S3}), we obtain the normalised joint probability distribution functions:
\begin{equation}
        \frac{P_{\textbf{p},\textbf{p}^\prime}}{\bar{n}^2}=\frac{P_{\textbf{q},\textbf{q}^\prime}}{\bar{n}^2} = 1+\frac{1}{\mu^2}\sin^2(\frac{\phi_L+\phi_R}{2}),\quad \text{and}\quad
        \frac{P_{\textbf{p},\textbf{q}^\prime}}{\bar{n}^2}=\frac{P_{\textbf{q},\textbf{p}^\prime}}{\bar{n}^2} = 1+\frac{1}{\mu^2}\cos^2(\frac{\phi_L+\phi_R}{2}).
        \label{S8}
\end{equation}

We emphasize that the interference between the scattered pairs, represented by the sine and cosine terms in the joint probability distribution functions mentioned above, is influenced by the global phase $\Phi = \phi_L + \phi_R$, rather than by the relative phase difference $\Phi = \phi_L - \phi_R$. This distinction arises from the slightly different, yet physically equivalent, geometry of our Rarity-Tapster interferometric scheme \cite{thomasMatterwave2022,rarityExperimental1990}. %

%-----------------------------------------------------------------------
\section{\label{Nonlocality} Bell correlation and nonlocality criterion}
From Eq.\,(5) of the main text and Eq.\,\eqref{S8}, we find that the Bell correlation function  $E(\phi_L,\phi_R)$ has the form,
\begin{equation}
    E(\Phi)\equiv E(\phi_L,\phi_R)=-A\cos(\phi_L+\phi_R),
    \label{S9}
\end{equation}
where the amplitude $A=(1+\bar{n})/(1+3\bar{n})$ is dependent on the average mode occupancy $\bar{n}$ of the halo \cite{hodgmanSolving2017}. Using the Bell correlation function, we can construct the CHSH-Bell parameter $S_{\mathrm{CHSH}}$ \cite{clauserProposed1969,rarityExperimental1990} for our experiment protocol using four pairs of phase settings, 
\begin{equation}
S_{\mathrm{CHSH}}=|E(\phi_L,\phi_R)+E(\phi_L^\prime,\phi_R)+E(\phi_L^\prime,\phi_R^\prime)-E(\phi_L,\phi_R^\prime)|,
\end{equation}
where any local hidden variable (LHV) theory must satisfy the bound $S_{\mathrm{CHSH}}\leq2$ independent of phase settings $\phi_L,\phi_L^\prime,\phi_R$, and $\phi_R^\prime$. As predicted by quantum theory, a violation of this CHSH-Bell inequality bound is possible through an appropriate selection of four pairs of phase settings - $(\phi_L,\phi_L^\prime,\phi_R,\phi_R^\prime)=(0,\pi/2,-\pi/4,-3\pi/4)$. These optimal settings differ from the typical choices \cite{rarityExperimental1990} by a sign resulting from our interferometric geometry and dependence on the global phase as opposed to the relative phase as seen in Refs.~\cite{rarityExperimental1990,lewis-swanProposal2015}. The amplitude range of $E$ limits the maximum achievable violation of the CHSH-Bell inequality for the chosen phase settings. Using the optimal phase settings, quantum mechanics leads us to expect a maximum value of $S_{\mathrm{CHSH}}$ to be $2\sqrt{2}A$ \cite{Clauser_Shimony_review_1978,lewis-swanProposal2015}, as indicated by the expression of $E(\Phi)$ in Eq.\,\eqref{S9}, where a minimum value of $A=1/\sqrt{2}$, is required to achieve a violation with $S_{\mathrm{CHSH}}>2$, while a maximum value of $A=1$ corresponds to $S_{\mathrm{CHSH}}=2\sqrt{2}$ \cite{lewis-swanProposal2015}. Due to the dependence of $A$ on the mode occupancy $\bar{n}$, we establish an upper bound for $\bar{n}\lesssim0.26$ in order to achieve $S_{\mathrm{CHSH}}>2$.
\begin{figure}
    \centering
    \includegraphics[scale=0.5]{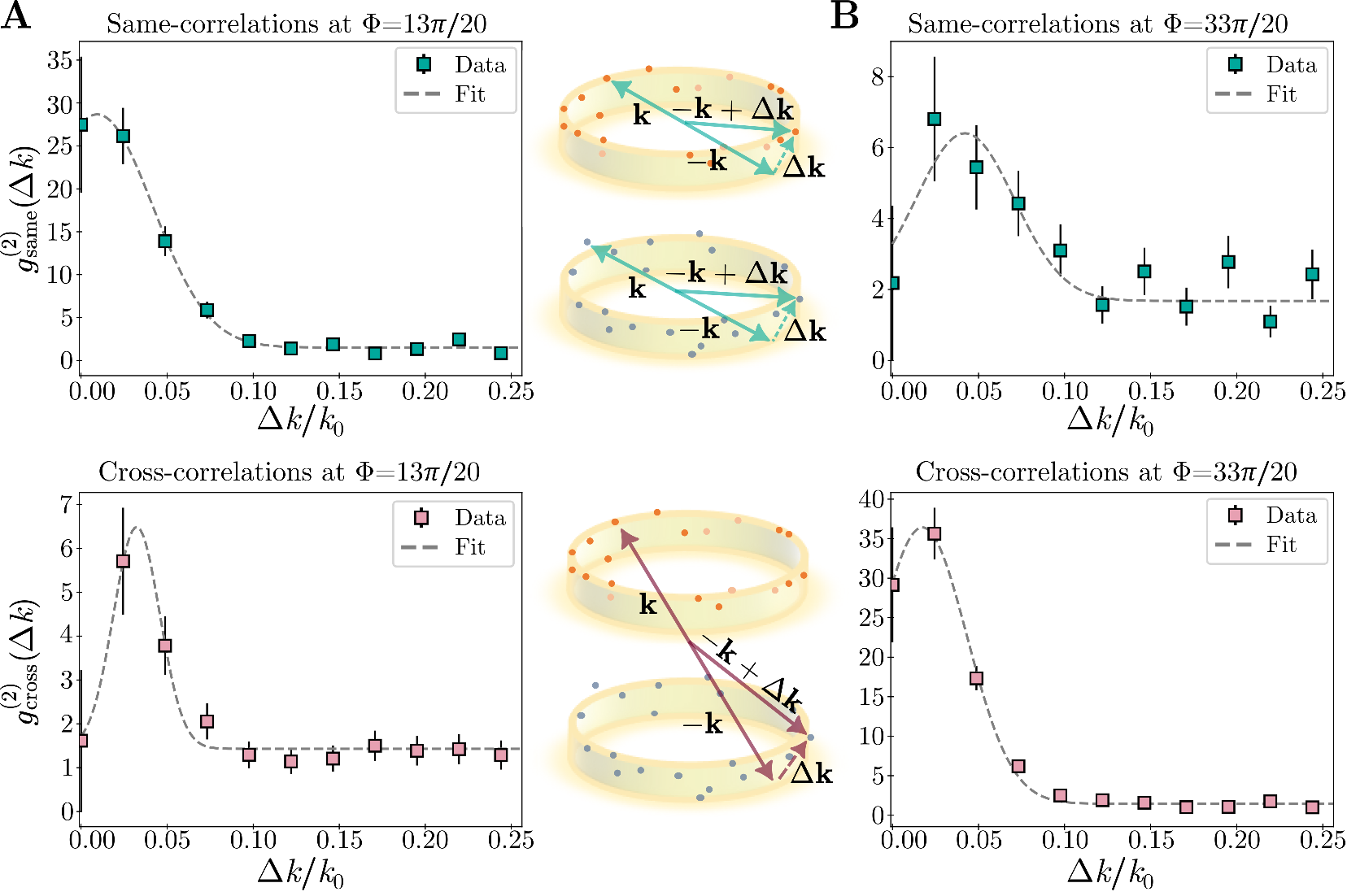}
    \caption{\textbf{Pair-correlations at interferometer output}. \textbf{(A)} Pair-correlation function $g^{(2)}(\Delta k)$ measured at the output momentum ports of the interferometer for global phase setting $\Phi=13\pi/20$. The two plots display the `same-correlations' (above) and `cross-correlations' (below) when measured within and across the two halos, respectively. \textbf{(B)} Pair-correlation function $g^{(2)}(\Delta k)$ measured at the interferometer output ports for $\Phi=33\pi/20$. From the two sets of plots at complimentary phase settings, we observe a shift in the correlation amplitudes ($g^{(2)}(0)$) between the `same-' and `cross-correlations' suggestive of particle-pair interference and entanglement of the initial state.}
    \label{fig:supplfig3}
\end{figure}

To confirm the absence of systematic biases in computing the Bell correlation function at each global phase setting used in Fig.~3B, we construct the Bell correlation function for momentum mode combinations that do not undergo two-particle interference under our scheme. Here, we ascribe the particular choice of combination 
of the momentum mode quartet based on their momentum vector difference $\Delta\textbf{k}$ (in the COM frame), where perfectly correlated pairs of modes ($\textbf{p},\textbf{p}^\prime,\textbf{q},\textbf{q}^\prime$) having $|\Delta\textbf{k}|\equiv \Delta k=0$ experience maximal interference. In contrast, pairs of atoms in momentum modes with a difference of $\Delta k>0$ deviate from maximal correlation strength and experience reduced interference at the beamsplitter. Figures \ref{fig:supplfig3}, A and B, display the pair-correlation amplitudes $g^{(2)}$ as a function of pair-mode momentum vector difference $\Delta\text{k}$ measured at the output of the interferometer for two global phase settings: (A) $\Phi=13\pi/20$  and (B) $\Phi=33\pi/20$. The `same-correlation' and `cross-correlation' plots represent measurements made within each halo and across the two halos, respectively. From these plots, with measurements made at each global phase setting, we can compute the Bell correlation function at a chosen $\Delta k$ using,
\begin{equation}
    E(\Phi)=\frac{g_{\mathrm{same},\Phi}^{(2)}(\Delta{k})-g_{\mathrm{cross},\Phi}^{(2)}(\Delta{k})}{g_{\mathrm{same},\Phi}^{(2)}(\Delta{k})+g_{\mathrm{cross},\Phi}^{(2)}(\Delta{k})}.
\end{equation}
We select $\Delta{k}\approx0.2$\texttimes${k}_0$ to measure the Bell correlation for uncorrelated atom pairs that do not experience two-particle interference in our scheme. The resulting values of $E$ for this combination of modes are displayed in Fig.~3B of the main text, showing no phase-dependent correlations, with values centred around the expected value of $E=0$. 

The observed shift in the correlation amplitudes $g^{(2)}(0)$ between the `same-' and `cross-correlation' plots in Fig.~\ref{fig:supplfig3}, A and B, is indicative of particle-pair interference in the system and entanglement of the initial state. We can express the pair-correlation function as $g^{(2)}(\Delta \textbf{k})=P_{\textbf{k},\textbf{k}^\prime}/\langle \hat{n}_\textbf{k} \rangle\langle \hat{n}_{\textbf{k}^\prime} \rangle$, and denote `same-correlations' ($g^{(2)}_{\mathrm{same}}$) with $(\textbf{k},\textbf{k}^\prime) \in \{(\textbf{p},\textbf{p}^\prime),(\textbf{q},\textbf{q}^\prime)\}$ and `cross-correlations' ($g^{(2)}_{\mathrm{cross}}$) with $(\textbf{k},\textbf{k}^\prime) \in \{(\textbf{p},\textbf{q}^\prime),(\textbf{q},\textbf{p}^\prime)\}$. As demonstrated in Eq.\,\eqref{suppl_eq:joint1}, the joint probability distribution functions are represented by sine and cosine terms dependent on the global phase $\Phi$, where the dependence arises from particle-pair interference of the initial Bell state (Eq.\,(1)) through the RT interferometer.

%-----------------------------------------------------
\subsection*{\textit{Nonlocality criterion}}
A requirement of the CHSH-Bell inequality violation is the implementation of independent phase settings between separated subsystems $L$ and $R$ corresponding to each atom in a scattered pair. In our current experiment protocol, due to the phase-imparting Bragg beams being significantly large in comparison to the scattering halo, all atoms acquire identical phase shifts resulting in equal phase settings between the two subsystems, i.e., $\phi_L=\phi_R$. This prevents us from demonstrating a violation of the CHSH-Bell inequality. 
To characterise the non-classical, non-local behaviour that we are able to demonstrate in our system, we instead arrive at a distinct nonlocality criterion that still enables us to rule out a large class of LHV theories. The derivation and use of this criterion follow from a similar demonstration by Shin \emph{et al.} (2019)\cite{shinBell2019d} for spin-entangled separated atom pairs.\\
Consider two separated subsystems $L$ and $R$, where quantities $\mathcal{L}$ and $\mathcal{R}$ are measured in each subsystem, respectively. The joint probability $P(\mathcal{L},\mathcal{R})$ for observing $\mathcal{L}$ and $\mathcal{R}$ fulfills the postulates of local realism if
\begin{equation}
    P(\mathcal{L},\mathcal{R})=\sum_\lambda p(\lambda)P(\mathcal{L}|\lambda)P(\mathcal{R}|\lambda),
\end{equation}
where $P(\mathcal{L}|\lambda)$ and $P(\mathcal{R}|\lambda)$ are the conditional probabilities of observing $\mathcal{L}$ or $\mathcal{R}$ given a value of some hidden variable $\lambda$, governed by the probability distribution $p(\lambda)$. Given some result $\mathcal{L}$, the conditional probability for observing $\mathcal{R}$ is
\begin{align}
    P(\mathcal{R}|\mathcal{L})&=\frac{P(\mathcal{L},\mathcal{R})}{P(\mathcal{L})}\notag\\
    &=\frac{\sum_\lambda p(\lambda)P(\mathcal{L}|\lambda)P(\mathcal{R}|\lambda)}{P(\mathcal{L})}\notag\\
    & = \sum_\lambda \frac{p(\lambda)P(\mathcal{L}|\lambda)}{P(\mathcal{L})}P(\mathcal{R}|\lambda)\notag\\
    &=\sum_\lambda P(\lambda|\mathcal{L})P(\mathcal{R}|\lambda).
\end{align}
Let us assume the quantity measured in $L$ to be $N_{i}^L$ produces binary outcomes for two local settings $i=\phi,\phi^\prime$, i.e., $N_{\phi}^L=\pm1$ and $N_{\phi^\prime}^L=\pm 1$, while, on average, the outcomes of the quantity $N_{i}^R$ measured in $R$ are assumed to be components of a vector of length 1. This implies that
\begin{equation}
    -1\leq\langle N_{i}^R\rangle \leq 1.
\end{equation}
We note here that the average outcomes in $R$ can be binary as well and for the following formalism and derivation, we do not specify any particular probability model in $R$ when calculating this average. Given the result $\mathcal{L}$, the average outcome in $R$ for the given setting, say $\phi$, is
\begin{align}
    \langle N_{\phi}^R \rangle_\mathcal{L} &= \sum_\mathcal{R} P(\mathcal{R}|\mathcal{L})N_{\phi}^R\notag\\
    &=\sum_\lambda P(\lambda|\mathcal{L})\sum_\mathcal{R} P(\mathcal{R}|\lambda)N_{\phi}^R, 
\end{align}
where
\begin{equation}
    -N_{\lambda}^R\leq\sum_\mathcal{R}P(\mathcal{R}|\lambda)N_{\phi}^R \leq N_{\lambda}^R
\end{equation}
and $N_{\lambda}^R$ is the length of the vector in $R$ given the value $\lambda$. This results in the upper bound
\begin{equation}
    \langle N_{\phi}^R\rangle_\mathcal{L} = \sum_\mathcal{R} P(\mathcal{R}|\mathcal{L})N_{\phi}^R \leq \sum_\lambda P(\lambda|\mathcal{L})N_{\lambda}^R = N_\mathcal{L}^R.
    \label{S16}
\end{equation}
Following our assumption that the averaged results of the quantity $N_\phi^R$ behave like components of a vector, we consider two quantities measured at orthogonal/complimentary settings, $\frac{1}{\sqrt{2}}(N_1^R-N_2^R)$, to express the average results of $N_\phi^R$. Thus, following the above result (Eq.\,\eqref{S16}) we obtain,
\begin{equation}
    -N_\mathcal{L}^R \leq \frac{\langle N_1^R\rangle_\mathcal{L} - \langle N_2^R\rangle_\mathcal{L}}{\sqrt{2}}\leq N_\mathcal{L}^R.
\end{equation}
We then multiply the two averages by the corresponding results in $L$. Since the outcomes in $L$ are taken to be binary, we obtain
\begin{equation}
    -N_\mathcal{L}^R \leq \frac{N_1^L\langle N_1^R\rangle_\mathcal{L} - N_2^L\langle N_2^R\rangle_\mathcal{L}}{\sqrt{2}}\leq N_\mathcal{L}^R.
    \label{S18}
\end{equation}
Finally, by averaging the inequality (Eq.\,\eqref{S18}) by the outcomes in $L$, we have
\begin{align}
    \sum_\mathcal{L} P(\mathcal{L})N_i^L\langle N_i^R \rangle_\mathcal{L} &= \sum_\mathcal{L} P(\mathcal{L})N_i^L \sum_\mathcal{R} P(\mathcal{R}|\mathcal{L})N_i^R\notag\\
    &=\sum_{\mathcal{L},\mathcal{R}}P(\mathcal{L},\mathcal{R})N_i^L N_i^R=\langle N_i^L N_i^R\rangle,
\end{align}
while $\sum_\mathcal{L} P(\mathcal{L})N_\mathcal{L}^R=\langle N^R\rangle \leq1$. Thus, the resulting inequality is
\begin{equation}
    |\langle N_1^LN_1^R \rangle - \langle N_2^LN_2^R\rangle|\leq\sqrt{2}.
    \label{S20}
\end{equation}

This inequality holds true for a wide range of LHV theories, where classical-like binary outcomes are observed in one or both subsystems and it is assumed that the averaged results in one of the subsystems behave like components of a vector (quantum-like), while no assumptions are made regarding the behaviour of the other subsystem \cite{wisemanSteering2007a,cavalcantiAnalog2015,cavalcantiExperimental2009a}.\\
In our experiment protocol, we define the measurement quantity $N_i=\hat{n}_{\textbf{k}_\mathrm{up}}-\hat{n}_{\textbf{k}_\mathrm{down}}$ where $i$ represents the phase setting ($\phi_L$ or $\phi_R$) set by the beamplitter pulse in the interferometer, $\hat{n}$ is the number operator acting on the momentum mode, while $\textbf{k}_\mathrm{up}\in\{\textbf{p},\textbf{p}^\prime\}$ and $\textbf{k}_\mathrm{down}\in\{\textbf{q},\textbf{q}^\prime\}$. 
We rewrite the inequality (Eq.\,\eqref{S20}) to form our nonlocality criterion
\begin{equation}
    \mathcal{C}(\phi,\phi+\pi/2)=|\langle N_\phi^LN_\phi^R \rangle - \langle N_{\phi+\pi/2}^LN_{\phi+\pi/2}^R\rangle|\leq\sqrt{2}.
    \label{S21}
\end{equation}
To test this inequality, we measure the Bell correlation function 
\begin{align}
    E(\Phi)\equiv E(\phi_L,\phi_R)&=\frac{\langle \hat{n}^{(L)}_\textbf{p}\hat{n}^{(R)}_{\textbf{p}^\prime}\rangle_{\phi_L,\phi_R} + \langle \hat{n}^{(L)}_\textbf{q}\hat{n}^{(R)}_{\textbf{q}^\prime}\rangle_{\phi_L,\phi_R} -\langle \hat{n}^{(L)}_\textbf{p}\hat{n}^{(R)}_{\textbf{q}^\prime}\rangle_{\phi_L,\phi_R} -\langle \hat{n}^{(L)}_\textbf{q}\hat{n}^{(R)}_{\textbf{p}^\prime}\rangle_{\phi_L,\phi_R}}{\langle \hat{n}^{(L)}_\textbf{p}\hat{n}^{(R)}_{\textbf{p}^\prime}\rangle_{\phi_L,\phi_R}+\langle \hat{n}^{(L)}_\textbf{q}\hat{n}^{(R)}_{\textbf{q}^\prime}\rangle_{\phi_L,\phi_R}+\langle \hat{n}^{(L)}_\textbf{p}\hat{n}^{(R)}_{\textbf{q}^\prime}\rangle_{\phi_L,\phi_R}+\langle \hat{n}^{(L)}_\textbf{q}\hat{n}^{(R)}_{\textbf{p}^\prime}\rangle_{\phi_L,\phi_R}}\notag\\
    &=\frac{\langle (\hat{n}^{(L)}_\textbf{p}-\hat{n}^{(L)}_\textbf{q})(\hat{n}^{(R)}_{\textbf{p}^\prime}-\hat{n}^{(R)}_{\textbf{q}^\prime})\rangle_{\phi_L,\phi_R}}{\langle (\hat{n}^{(L)}_\textbf{p}+\hat{n}^{(L)}_\textbf{q})(\hat{n}^{(R)}_{\textbf{p}^\prime}+\hat{n}^{(R)}_{\textbf{q}^\prime})\rangle_{\phi_L,\phi_R}}\notag\\
    &=\frac{\langle N_{\phi_L}^LN_{\phi_R}^R\rangle}{\langle (\hat{n}^{(L)}_\textbf{p}+\hat{n}^{(L)}_\textbf{q})(\hat{n}^{(R)}_{\textbf{p}^\prime}+\hat{n}^{(R)}_{\textbf{q}^\prime})\rangle_{\phi_L,\phi_R}}.
    \label{S22}
\end{align}
When the system is composed of a single scattering pair, the denominator in Eq.\,\eqref{S22} is equal to 1. Thus, we obtain a simplified expression for $E(\phi_L,\phi_R)=\langle N_{\phi_L}^LN_{\phi_R}^R \rangle$ and transform the nonlocality criterion (Eq.\,\eqref{S21}) to obtain,
\begin{equation}
    \mathcal{C}(\phi,\phi+\pi/2) = |E(\phi,\phi)-E(\phi+\pi/2,\phi+\pi/2)|\leq \sqrt{2}.
\end{equation}
 An equivalent of this expression is seen in Eq.\,(7), expressing the above inequality in terms of the global phase parameter $\Phi=\phi_L+\phi_R$. 

%%------------------------------------------------------------------------
% ************** End of supplementary ******************

\end{document}